\documentclass[12pt]{article}
\usepackage{fixltx2e}
\usepackage{upgreek}
\usepackage{amssymb,amsmath}
\usepackage[]{breqn}
\usepackage{array}
\usepackage{float}
\usepackage{graphicx}
\usepackage{booktabs}
\usepackage{tabularx}
 
\def\'#1{{\accent19\ifx #1i \i\else #1\fi}}

\def\be{\begin{equation}}
\def\ee{\end{equation}}
\def\bea{\begin{eqnarray}}
\def\eea{\end{eqnarray}}

\MakeRobust{\overleftarrow}
\renewcommand\[{\begin{dmath}}
\renewcommand\]{\end{dmath}}

\newcommand{\boldmathtau}{\mbox{\boldmath$\tau$\unboldmath}}
\newcommand{\boldmathphi}{\mbox{\boldmath$\phi$\unboldmath}}

\newcommand{\boldmathPsi}{\mbox{\boldmath$\Psi$\unboldmath}}

\newbox\Ancha
\catcode`@=11
\newdimen\ex@
\title{ Heavy quarks within the electroweak multiplet }

\author{J. Besprosvany and R. Romero}
\date{Instituto de F\'{\i}sica, Universidad Nacional Aut\'onoma de M\'exico,
Apartado Postal 20-364,   Ciudad de M\'exico 01000, M\'exico }

\begin{document}

\maketitle









\jot = 1.5ex
 \def\baselinestretch{1.9}
\parskip 5pt plus 1pt

\begin{abstract}
Standard-model fields and  their  associated
electroweak   Lagrangian  are equivalently expressed in a shared spin basis.  The  scalar-vector terms are written with scalar-operator  components acting on
 quark-doublet elements, and shown  to be  parametrization-invariant.
       Such terms, and the  t- and  b-quark Yukawa  terms are linked by    the identification  of the common   mass-generating    Higgs  operating upon the other fields, after acquiring a vacuum expectation value
$v$. 
    Thus,  the customary vector  masses   are related to the fermions',
fixing  the t-quark mass $m_t$  with the relation $m^2_t+m^2_b=v^2/2$
either for maximal hierarchy, or given the b-quark mass $m_b$,  implying
$m_t
\simeq 173.9$ GeV, for
$v=246$ GeV.  A sum  rule is derived  for  all quark masses that generalizes this restriction.
An interpretation follows that electroweak bosons and heavy quarks belong in a multiplet.
\end{abstract}

\vskip 2cm



\vskip 1cm
Keywords: Top quark, mass, multiplet,   Lagrangian, spin, electroweak

 \baselineskip 22pt\vfil\eject \noindent

\section{Introduction}



The standard model (SM)  describes elementary-particle features and  their interactions, which is praiseworthy, given its relatively limited required  input,
 consisting of specific gauge and flavor symmetries, representations,  and parameters, yet
aspects remain within the model whose origin and connection to other tenets is absent, and that need to be addressed.


 Thus, among its successes,   the SM   predicts  mass  values for the W and Z  bosons\cite{Glashow} that  carry  the short-range electroweak interaction, in terms of electroweak parameters,  through the    Higgs mechanism\cite{HiggsMech,HiggsMechHiggs}.
 However, one salient SM problem  is that the    fermion sector and its  masses  remain  arbitrary, as they arise  from  Lagrangian terms, independent from the boson  elements.


 The electroweak sector hints it  may provide this    link, given that  the  W and Z  vectors have  universal couplings to SM fermions, and  the Higgs field  collectively   gives mass to fermions and bosons.
In addition, the similar order of magnitude of the  measured masses\cite{ParticleData}   of  the W, Z,     the recently discovered  scalar  excitation, associated  with
the Higgs\cite{HiggsIndi}, and the top quark (with the bottom quark's  the next highest),
 suggests   connections among them, and thus, a  common  energy  scale.
Furthermore,  fermions occupy the  spin-1/2 and fundamental representations of  the Lorentz and  scalar groups,
   respectively,   as  vector bosons  belong to the    adjoint representation of each group,\footnote{As the Higgs occupies the SU$_L$(2) fundamental representation.}$^,$\footnote{For the Abelian  hypercharge group U(1)$_Y$, gauge invariance ensures boson-fermion  quantum-number  additivity.}   which implies  bosons   can be constructed in terms of fermions,
 suggesting  composite structures and/or a  common  origin.

The above motivates looking for   a formalism that  takes account of discrete degrees of freedom  in a single basis,  including   group representation properties, such as
  the fermion-boson fundamental-adjoint duality for   the Lorentz-scalar  representations, and  that    describes the combined action of operators on fields.

A previously proposed SM extension\cite{BesproIni},   based on a shared  extended spin space, with a matrix formalism,  satisfies these requirements, as it replicates      SM fields with their  features, and matrix multiplication accounts for operator action on fields.
  This  space contains a  (3+1)-dimensional [d] subspace  and   one  beyond 3+1, linked,    respectively, to Lorentz and   scalar degrees of freedom\cite{BesproRicardo}. At each dimension, a finite number of Lorentz-invariant partitions are generated with specific symmetries and representations, reproducing  particular SM features, where the   cases with dimension 5+1\cite{Besprosub}, 7+1\cite{Romero}, and 9+1\cite{bespro9m1}   were studied.



In this connection, it is worth recalling that a basis or representation choice  can be useful, even essential,
 in the description of a   system and its dynamics. It may reveal otherwise-hidden connections between its components,
and provide a  simpler framework to  understand   physical properties. Such a basis may  describe effective degrees of freedom\cite{Landau} accounting for collective interactions, allowing for a simpler near free-particle description, in a first approximation.   For example,   nucleon  and associated boson  interactive configurations give a tractable account of nuclear-motion modes\cite {nuclear}. Within condensed matter and  low-temperature superconducting systems, a residual attractive interaction related to   phonons  couple electrons into Cooper pairs\cite{Schrieffer}, which    propagate freely, and
  lead to frictionless currents.   In an application  of this theory to  quantum field theory and elementary particles\cite{NambuJona},
 a  four-fermion   interaction   produces fermion     and composite-boson   masses,
 linking  their values.
    The quark model\cite{quarkmodel} conceives mesons and baryons in terms of constituent (dressed) quarks.




  Leaving aside  the more speculative nature of the spin SM extension, but  complementarily  to it,  in this paper, we use it as a basis  to derive   SM connections,   and the fields'  mass values in particular:
       SM  heavy-fermion ($F$),  vector ($V$), and scalar ($S$)  fields are equivalently expressed in terms of  the   obtained common  basis\cite {BesproIni}
       for  both Lorentz and electroweak  degrees of freedom,  in turn,
        recasting  their   Lagrangian   components  ${\cal L}={\cal L}_{FV}+ {\cal L}_{SV}+ {\cal L}_{SF}$; the identification of the  scalar operator  within  the
     ${\cal L}_{SF}$ and  ${\cal L}_{SV}$ vertices links univocally its defining (mass) parameters.
   Indeed, such universal electroweakly-invariant terms    lead,
   under the Higgs mechanism, to  a     scalar  whose lowest-energy condensate state pervades space,
  and generates  particle  masses through its vacuum expectation value $v$.     Within the  spin basis, this mechanism is similarly  represented;
   as these fields shape elements on a matrix space, with  a  single associated    scalar    operator acting upon the others, their  mass-generation  property
    relates  their coefficients.

Next, as we give the paper's organization,  we sketch the argument in more detail.   Section  2  reviews the applied spin-extended space for symmetry generators  and states.  The paper focuses  then on  the (7+1)-d case  that can describe the electroweak
sector, and a quark doublet. For all sectors, ${\cal L}_{FV}$, ${\cal L}_{SV}$,  $ {\cal L}_{SF}$,
  the  conventional and spin-space Lagrangian  are equivalent, which is shown  term-by-term  in Appendices 1,2.
 Section 3  chooses one among two     vector bases    within ${\cal L}_{FV}$, where vectors with chiral properties are adequate.    Section 4  writes
  ${\cal L}_{SV}$  equivalently  with    combinations  of the scalars   and  their conjugates,  with universal  couplings to  vectors, shown explicitly in Appendix 2;
  similarly for the spin-base representation, in which these two  scalars    induce  a projection  to flavor-doublet components (as t,b  quarks).
Schematically,   given the  spin-space basis element $B_f$ for a field $f(x)$, we  write ${\cal L}_{SV}$    in terms of  $B_{S'}$ containing these two scalar components, obtaining the vector  mass   squared within
    ${\cal L}_{FV}$     as  $[B_{S'},B_V]^\dagger [B_{S'},B_V]$.
      In Section 5, we show
      that  the fermion masses     within  ${\cal L}_{SF}$ can be written $[B_S,B_F] $, where $B_F$ contains two terms  with appropriate  Yukawa coefficients.
     Within the   spin-basis formalism,  we derive that  $B_{S'}$,         $B_{S}$   have the same operator structure;  given their mass-giving nature, the  identification of  these operators  and their coefficients  translates a $v$-normalization restriction   on   $B_{S'}$
to  $B_{S}$,      implying a  relation for the t and b  quark masses. Section 6  shows a  procedure exists that generalizes consistently
 this relation to all quarks in terms of a sum rule for their masses,  taking advantage of the chiral projection properties of the scalar field in the spin basis.
      In   Section 7,  we draw conclusions.
       We work in  the classical framework afforded by the Lagrangian,  and  at tree-level, but  also rely on a quantum-mechanical interpretation.








\section{Symmetry generators and states in spin-extended space }


In the following, we introduce  the spin basis  and its  main features, where more information may be found in previous
treatments\cite{BesproRicardo}-\cite{bespro9m1}.
Mainly, it describes SM discrete degrees of freedom in a  single scheme, namely,  for the Lorentz    and scalar groups,   and for   both symmetry generators   and state  representations,
using a common matrix space:

\subsection{Matrix space}

Such a  space is
rendered by a   Clifford algebra $\mathcal{C}_{N}$,
  generated by a set of   even-$N$   $2^{N/2}\times 2^{N/2}$   gamma  matrices,
  obeying the defining property\cite{Snygg}

\begin{equation}
\gamma_{\alpha}\gamma_{\beta}+\gamma\emph{}_{\beta}\gamma_{\alpha}=2g_{\alpha\beta},\label{eq:1}
\end{equation}

\noindent where $g_{\alpha\beta}$ is the metric tensor with signature $(+,-,...,-)$
and\footnote{Following standard practice, the label 4 is omitted.} $\alpha,\beta=0,1,\ldots3,5,\ldots,N$,
 whose combinations produce a complex matrix-space  with  dimension $2^{N}$.

The gamma matrices have      Hermiticity  properties

\begin{equation}
\begin{array}{rc}
\gamma_{0}^{\dagger}=\gamma_{0},\\
\gamma_{  \delta}^{\dagger}=-\gamma_{ \delta} &   \delta= 1,\ldots3,5,\ldots,N.
\end{array}\label{eq:1-1}
\end{equation}

\subsection{Operators and symmetry transformations }

\noindent
The   Lorentz generators and  transformations acting
on spinors have standard expressions in the 4-d Clifford
algebra $\mathcal{C}_{4}$, namely,

\begin{equation}
\label{sigmamunugen}
\begin{array}{ccc}
\sigma_{\mu\nu}=\frac{i}{2}\left[\gamma_{\mu},\gamma_{\nu}\right] & \textrm{with} & \mu,\nu=0,\ldots,3,\end{array}
\end{equation}

\noindent
\begin{equation}
S(\Lambda)=e^{-\frac{i}{4}\sigma_{\mu\nu}\omega^{\mu\nu}},\label{eq:3}
\end{equation}
with the $(3+1)$-d  gamma matrices $\gamma_{\mu}$ transforming
as vectors,
  while the remaining $N-4$ gamma
matrices $\gamma_{a}$, $a=5,\ldots,N$, and their products commuting
with $\sigma_{\mu\nu}$,
so they  are indeed Lorentz scalars
  identified with generators of continuous
symmetries, either gauge or global.
Together with the $4$-d pseudoscalar \noindent
\begin{equation}
\tilde{\gamma}_{5}\equiv-i\gamma_{0}\gamma_{1}\gamma_{2}\gamma_{3},\label{eq:4-1}
\end{equation}
 the scalars are accommodated in the unitary symmetry set
\begin{equation}
\label{SNm4}
\mathcal{S}_{N-4}=\frac{1}{2}(1+\tilde{\gamma}_{5})\textrm{U}\left(2^{(N-4)/2}\right)\oplus\frac{1}{2}(1-\tilde{\gamma}_{5})\textrm{U}\left(2^{(N-4)/2}\right),
\end{equation}
 where $1$ stands for the $N$-d identity matrix.


A  projector operator $\mathcal{P}$,
obtained from elements of $ \mathcal{S}_{N-4}$,  within a limited number partitions, is  chosen  to fit as closely the SM. The combined operator that  acts
on both the Lorentz generators $\mathcal{J}_{\mu\nu}=i\left(x_{\mu}\partial_{\nu}-x_{\nu}\partial_{\mu}\right)+\frac{1}{2}\sigma_{\mu\nu}$
and the $\mathcal{S}_{N-4}$ symmetry-operator space
is likewise projected
\begin{equation}
\begin{array}{c}
\mathcal{J}'_{\mu\nu}=\mathcal{P}\mathcal{J}_{\mu\nu}=\mathcal{P}\left[i\left(x_{\mu}\partial_{\nu}-x_{\nu}\partial_{\mu}\right)+\frac{1}{2}\sigma_{\mu\nu}\right],\\
\\
 \mathcal{S}'_{N-4}=\mathcal{P}\mathcal{S}_{N-4}.
\end{array}  \label{Project}
\end{equation}
Lorentz  transformations are thus
\noindent
\begin{equation}
S(\Lambda)=e^{-\frac{i}{4}{\mathcal{P}}\omega^{\mu\nu}\sigma_{\mu\nu}}.\label{LorentzTransfo}
\end{equation}
and
scalar transformations  have the  form
\begin{equation}
U=\exp\left[-iI_{a}\alpha_{a}(x)\right],\label{ScalarTransfo}
\end{equation}
 with $I_{a} \in  \mathcal{S}'_{N-4} $.
Symmetry generators within this space are described schematically in Fig. 1  in Ref. \cite{Romero}.

The inner product of two fields is defined according to a matrix space
\begin{equation}
\left\langle \phi\mid\Psi\right\rangle =\text{tr}\left(\phi{}^{\dagger}\Psi\right).\label{eq:6-1}
\end{equation}

 Under a unitary transformation,  $\Psi\rightarrow U \Psi U^\dagger$,
given the ket-bra matrix structure\cite{BesproRicardo}, with the bras  interpreted as conjugate states.
Thus,  a Hermitian operator $Op$ within this space  characterizes a state  $\Psi$ with the eigenvalue rule
\begin{equation}\label{eq:6}
[Op,\Psi]=\lambda  \Psi , \end{equation}
for real $\lambda$.
This definition is consistent with the action of a derivative operator on a Hilbert space:
 $[\overrightarrow{\partial},\Psi]=[-\overleftarrow{\partial},\Psi]=[\Psi, \overleftarrow{\partial}]$.
The direct product   tr$\Psi_b^\dagger\Psi_a$ is also  consistent associativity-wise with the operator rule,
as   ${\rm tr }[Op,\Psi_b]^\dagger \Psi _a={\rm tr } \Psi^\dagger_b[ Op^\dagger,\Psi_a ]. $

\subsection{Field Representation}

Fields are usually assumed to exist on a Cartesian basis;  for example, a vector field has components $A_\mu(x)=g_{\mu} \  ^\nu A_\nu(x)$; alternatively, in the
    spin  basis, it is expressed as
$A_\mu(x)(\gamma_0   \gamma^\mu ) _{\alpha \beta }$ (the  $\alpha \beta $ indices now specify the vector character.)

More generally, a physical field  with scalar quantum numbers is associated with elements of $\mathcal{C}_{N}$,
classified by operators from $\mathcal{C}_{4}\otimes\mathcal{S}_{N-4}$,
so it  has the structure

\begin{equation}
\left(\mbox{elements of 3+1 space }\right)\times\left(
\text{ elements of \ensuremath{\mathcal{S}_{N-4}}}
\right).\label{eq:10}
\end{equation}

Fig. 2 in Ref. \cite{Romero}  shows the corresponding  Lorentz states: scalars, vectors, fermions, and anti-symmetric tensors, arranged in the same matrix space.
 Next, we provide more details on the first
three (physical) fields.

\subsubsection*{Fermion field}

When  $\Psi$ is a spin-1/2 particle, it may be seen schematically    conformed as
 $\Psi\sim\left | \psi_1\right \rangle \left |a_1 F_1\right\rangle
 \left\langle F_2\right |$, with the ket  carrying spin-1/2    and gauge-group fundamental representation   $\psi_i$, $a_i$  quantum numbers, respectively, and both the bra and ket carrying  flavor group  $F_i$.


More specifically, a fermion can have the form
\begin{equation}
\psi_{\alpha}^{a}(x)L^{\alpha}P_{F}\Gamma_{a}^{F},\label{eq:16-1}
\end{equation}

\noindent where $\Gamma_{a}^{F}$ is an element of $ \mathcal{S}_{N-4} $,
and $L^{\alpha}$ represents a spin polarization component, e.g., $L^{1}=\left(\gamma_{1}+i\gamma_{2}\right)$.
The operator $P_{F}$ is a projection operator,  e. g.,    $P_{F}=L_5$, where
\begin{eqnarray} \label{projeRL}
  R_5=\frac{1}{2}(1+\tilde \gamma_5),  \    \  L_5=\frac{1}{2}(1-\tilde \gamma_5),
    \end{eqnarray}
implying
\begin{equation}
P_{F}\gamma^{\mu}=\gamma^{\mu}P_{F}^{c},\label{eq:17-1}
\end{equation}

\noindent with $P_{F}^{c}=1-P_{F}$,
so that Lorentz and
gauge generators act trivially on its rhs when evaluating commutators
as in Eq. (\ref{eq:6}), since $ P_F^{c}   P_F =(1- P_F ) P_F  =0$.

Thus, for $U$ accounting  for the Lorentz representation in Eq.  \ref{LorentzTransfo} and the scalar transformation in Eq. \ref{ScalarTransfo},   $\Psi$ transforms, unlike vector and scalar fields, as
\begin{equation}
\Psi\rightarrow U\Psi.\label{eq:12-1}
\end{equation} This
leads to fermions transforming as the fundamental representation of
both the Lorentz and gauge groups.

\subsubsection*{Vector field}
\noindent We may view vectors   constructed as
$\Psi\sim\left|\psi_1\right\rangle\left|a_1\right\rangle\left\langle a_2\right|\left\langle \psi_2\right|$, with the bra-ket configuration producing Lorentz  vector and gauge group adjoint   configurations, given
 the vector and scalar    $\gamma^{\mu}$, $\mu=0,\ldots,3$ and
$\gamma^{a}$, $a=5,\ldots,N$, respective transformation  properties. Thus a vector field has form
\begin{equation}
A_{\mu}^{a}(x)\gamma_{0}\gamma_{\mu}I_{a},\label{eq:14}
\end{equation}
\noindent where $\gamma_{0}\gamma_{\mu}\in\mathcal{C}_{4}$ and $I_{a}\in\mathcal{S}'_{N-4}$
is a generator of a given unitary group.

\subsubsection*{Scalar field}

$\Psi\sim\left|\psi_1\right\rangle\left|a_1\right\rangle\left\langle a_2\right|\left\langle \psi_2\right|$, with the bra-ket configuration producing Lorentz  vector and gauge group fundamental   configurations. In this case, a ket contains right-handed and a bra left-handed  spin-1/2 components (or vice versa),  reproducing the mass term and Higgs quantum numbers.

\begin{equation}
\phi^{a}(x)\gamma_{0}\Gamma_{a}^{S},\label{eq:15-1}
\end{equation}

\noindent with $\Gamma_{a}^{S}$ an element of $\mathcal{S}_{N-4}$.

\subsection{Lagrangian formulation}

Interactive Lagrangians\cite{BesproRicardo} can be given in
terms of vector, scalar and fermion fields conforming to the general
structure of   operator action as in Eq.  \ref{eq:6} and the inner product in Eq. \ref{eq:6-1}.
\noindent For example, a gauge-invariant fermion-vector Lagrangian
is given by
\begin{equation}
\dfrac{1}{N_{f}}\text{tr}\Psi^{\dagger}\left\{ \left[i\partial_{\mu}-gA_{\mu}^{a}(x)I_{a}\right]\gamma^{0}\gamma^{\mu}-M\gamma^{0}\right\} \Psi,\label{eq:18-1}
\end{equation}
\noindent where $\Psi$ is a fermion field as in Eq. (\ref{eq:16-1}), $g$ is the coupling constant, $M$ is an appropriate mass operator, and $N_{f}$ contains the normalization. In the next subsection, we address    the spin model  in 7+1  d   in connection with the SM, and whose basis states will allow
to write ${\cal L}_{FV}$, ${\cal L}_{SV}$,  $ {\cal L}_{SF}$ in the next Sections.

\subsection{(7+1)-dimensional model}

We next make a brief description of resulting states in a (7+1)-dimensional
spin space under a useful partition for the SM description,
 sketching  the way to obtain it, and providing   graphic description.

\subsubsection{Operators}

The Clifford algebra is generated by eight $16\times16$ matrices

\begin{equation}
\gamma_{0},\gamma_{1},\ldots,\gamma_{8}.\label{eq:2.5.1}
\end{equation}

\noindent The matrices $\gamma^{0}$, $\gamma^{i}$, $i=1,2,3$ correspond
to the Lorentz generators $\sigma_{\mu\nu}$, given in general in Eq. \ref{sigmamunugen} and the remaining four
matrices, together with all their different products, comprise
the set $ \mathcal{S}_{N-4}$ of scalars, with a cardinality of 32. This
set   is, from Eq. \ref{SNm4},   ${\textstyle \mathcal{S}_4=P_{+}\textrm{U}\left(4\right)\oplus P_{-}\textrm{U}\left(4\right)}$,
with $P_{\pm}=\frac{1}{2}\left(1\pm\tilde{\gamma}_{5}\right)$,
$1$ the $16\times16$ identity matrix and $\tilde{\gamma}_{5}$
the 4-d chirality matrix. The elements of $\textrm{U}\left(4\right)$
consist of four matrices $\gamma_{a}$, $a=5,\ldots,8$, six pairs
$\gamma_{ab}\equiv\gamma_{a}\gamma_{b}$, $a<b$, four triplets $\gamma_{abc}\equiv\gamma_{a}\gamma_{b}\gamma_{c}$,
and one quadruplet $\gamma_{5}\gamma_{6}\gamma_{7}\gamma_{8}$. The
Cartan subalgebra $\mathfrak{h}$ of $ \mathcal{S}_{N-4}$ contains eight
elements, and a suitable choice is given by

\begin{equation}
\begin{array}{cccccccc}
1, &\tilde{ {\gamma}}_{5}, & \gamma_{5}\gamma_{6}, & \gamma_{7}\gamma_{8}, & \gamma_{5}\gamma_{6}\gamma_{7}\gamma_{8}, & \gamma_{5}\gamma_{6}\tilde{\gamma}_{5}, & \gamma_{7}\gamma_{8}\tilde{\gamma}_{5}, & \gamma_{5}\gamma_{6}\gamma_{7}\gamma_{8}\tilde{\gamma}_{5}.\end{array}\label{eq:2.5.2}
\end{equation}

\begin{figure}[H]
\begin{centering}
\includegraphics[scale=0.7]{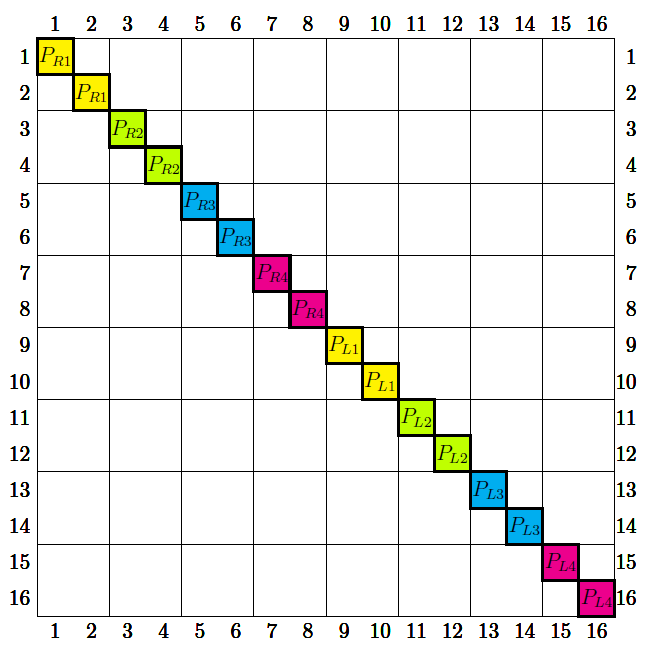}
\par\end{centering}
\caption{(Color online) Matrix representation of the Cartan basis (cf. eq.
(\ref{eq:2.5.2})) in extended spin space in 7+1 dimensions. The eight-dimensional
basis is represented here in terms of the projection operators $P_{R,Li}$,
$i=1,\ldots,4.$ The subscripts $R,L$ refer to the chirality: $R$
for operators containing $1+\tilde{\gamma}_{5}$ (right-handed), and
$L$ for operators containing $1-\tilde{\gamma}_{5}$ (left-handed).}   \label{projections}
\end{figure}

\noindent Since  $\mathfrak{h}$ is conformed of   all simultaneously diagonalizable operators, it is convenient to recast this basis in terms of
the projection operators

\begin{equation}
\begin{array}{cc}
P_{R1}= & \frac{1}{8}(1+\tilde{\gamma}_{5})(1+i\gamma_{5}\gamma_{6})(1+i\gamma_{7}\gamma_{8}),\\
P_{R2}= & \frac{1}{8}(1+\tilde{\gamma}_{5})(1+i\gamma_{5}\gamma_{6})(1-i\gamma_{7}\gamma_{8}),\\
P_{R3}= & \frac{1}{8}(1+\tilde{\gamma}_{5})(1-i\gamma_{5}\gamma_{6})(1+i\gamma_{7}\gamma_{8}),\\
P_{R4}= & \frac{1}{8}(1+\tilde{\gamma}_{5})(1-i\gamma_{5}\gamma_{6})(1-i\gamma_{7}\gamma_{8}),\\
P_{L1}= & \frac{1}{8}(1-\tilde{\gamma}_{5})(1+i\gamma_{5}\gamma_{6})(1+i\gamma_{7}\gamma_{8}),\\
P_{L2}= & \frac{1}{8}(1-\tilde{\gamma}_{5})(1+i\gamma_{5}\gamma_{6})(1-i\gamma_{7}\gamma_{8}),\\
P_{L3}= & \frac{1}{8}(1-\tilde{\gamma}_{5})(1-i\gamma_{5}\gamma_{6})(1+i\gamma_{7}\gamma_{8}),\\
P_{L4}= & \frac{1}{8}(1-\tilde{\gamma}_{5})(1-i\gamma_{5}\gamma_{6})(1-i\gamma_{7}\gamma_{8}),
\end{array}\label{eq:2.5.3}
\end{equation}

\noindent which run along the diagonal in the matrix space (Fig. \ref{projections}).

The operators that classify the states, with examples in terms of the projectors,   consist of the baryon-number
operator

\begin{equation}
B=\frac{1}{6}(1-i\gamma_{5}\gamma_{6})=\dfrac{1}{3}(P_{R3}+P_{R4}+P_{L3}+P_{L4}),\label{eq:2.5.4}
\end{equation}

\noindent the U(1) hypercharge generator

\begin{equation}
\begin{split}Y_o & =\frac{1}{3}\left(4P_{R3}-2P_{R4}+P_{L3}+P_{L4}\right),\\
 & =\frac{1}{6}\left(1-i\gamma_{5}\gamma_{6}\right)\left(1+i\frac{3}{2}(1+\tilde{\gamma}_{5})\gamma_{7}\gamma_{8}\right),
\end{split}
\label{eq:2.5.5}
\end{equation}

\noindent and   $I_{3}$ within  the SU(2) weak isospin generators

\begin{equation}
\begin{split}I_{1} & ={\displaystyle \frac{i}{8}(1-\tilde{\gamma}_{5})(1-i\gamma_{5}\gamma_{6})\gamma^{7}},\\
I_{2} & {\displaystyle =\frac{i}{8}(1-\tilde{\gamma}_{5})(1-i\gamma_{5}\gamma_{6})\gamma^{8}},\\
I_{3} & =\frac{1}{2}(P_{L3}-P_{L4})=\frac{i}{8}(1-\tilde{\gamma}_{5})(1-i\gamma_{5}\gamma_{6})\gamma_{7}\gamma_{8}.
\end{split}
\label{eq:2.5.6}
\end{equation}

\noindent The charge operator is defined in the standard way by the
Gell-Mann--Nishijima relation

\begin{equation}
Q=I_{3}+\frac{Y_o}{2}.\label{eq:2.5.7}
\end{equation}

There are also flavor operators, forming the groups $\textrm{SU}(2)_{f}$,
$\textrm{SU}(2)_{\hat{f}}$, $\textrm{U}(1)_{f}$, and $\textrm{U}(1)_{\hat{f}}$,
and given by

\begin{equation}
\begin{split}f_{1} & =\frac{i}{8}\left(1+\tilde{\gamma}_{5}\right)\left(1+i\gamma^{5}\gamma^{6}\right)\gamma^{7},\\
f_{2} & =\frac{i}{8}\left(1+\tilde{\gamma}_{5}\right)\left(1+i\gamma^{5}\gamma^{6}\right)\gamma^{8},\\
f_{3} & =\frac{i}{8}\left(1+\tilde{\gamma}_{5}\right)\left(1+i\gamma^{5}\gamma^{6}\right)\gamma^{7}\gamma^{8},
\end{split}
\label{eq:2.5.8}
\end{equation}

\begin{equation}
\begin{split}\hat{f}_{1} & =\frac{i}{8}\left(1-\tilde{\gamma}_{5}\right)\left(1+i\gamma^{5}\gamma^{6}\right)\gamma^{7},\\
\hat{f}_{2} & =\frac{i}{8}\left(1-\tilde{\gamma}_{5}\right)\left(1+i\gamma^{5}\gamma^{6}\right)\gamma^{8},\\
\hat{f}_{3} & =\frac{i}{8}\left(1-\tilde{\gamma}_{5}\right)\left(1+i\gamma^{5}\gamma^{6}\right)\gamma^{7}\gamma^{8},
\end{split}
\label{eq:2.5.9}
\end{equation}

\noindent respectively for $\textrm{SU}(2)_{f}$ and $\textrm{SU}(2)_{\hat{f}}$,
and

\begin{equation}
f_{0}=i\gamma^{5}\gamma^{6}\tilde{\gamma}_{5},\label{eq:2.5.10}
\end{equation}

\begin{equation}
\hat{f}_{0}=i\gamma^{5}\gamma^{6},\label{eq:2.5.11}
\end{equation}

\noindent for $\textrm{U}(1)_{f}$, and $\textrm{U}(1)_{\hat{f}}$.
The operators $f_{3}$, $\hat{f}_{3}$, $f_{0}$ and $\hat{f}_{0}$
belong to $\mathfrak{h}$. In Fig. \ref{projections} the matrix space is
represented schematically. The diagonal operators classify the states (off-diagonal)
acting from the left for states in the same row, and from the right
for states in the same column, which is consistent with matrix multiplication.

We also define a combination of diagonal flavor operators that further classifies states, given by

\begin{equation}
\hat{F}=-\dfrac{1}{4}\left(\hat{f}_{0}+4\hat{f}_{3}-8f_{3}\right). \label{eq: Fop}
\end{equation}

\begin{figure}[H]
\begin{centering}
\includegraphics[scale=0.7]{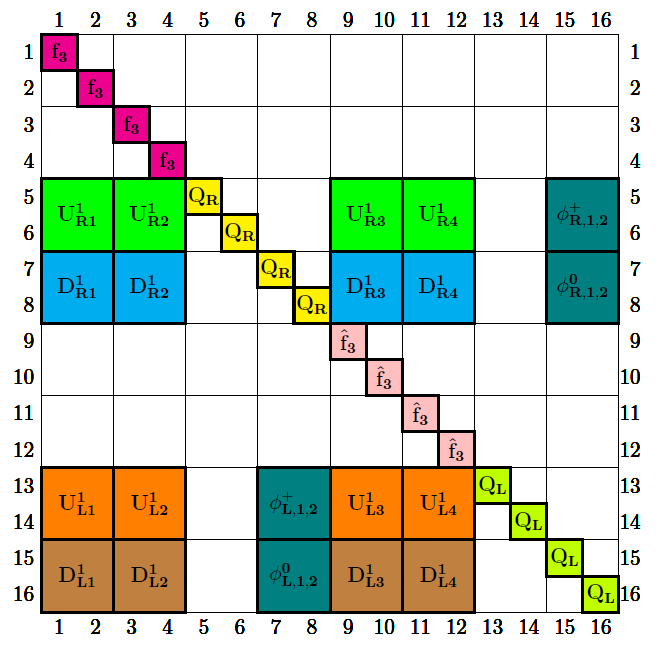}
\par\end{centering}
\caption{(Color online) Matrix representation of operators, massless quarks
$\left(U_{L,Ri}^{1},\,D_{L,Ri}^{1},\,i=1,\ldots,4\right)$ and Higgs
$\left(\phi_{1,2}^{+},\,\phi_{1,2}^{0}\right)$ degrees of freedom
in $\left(7+1\right)$-d spin space. The chiral projections of the
diagonal operators $B$, $I_{3}$ and $Y_o$ are grouped together and
represented by the sets ${\bf Q}_{R}=\frac{1}{2}\left(1+\tilde{\gamma}_{5}\right)\left(B,I_{3},Y_o\right)$
and ${\bf Q}_{L}=\frac{1}{2}\left(1-\tilde{\gamma}_{5}\right)\left(B,Y_o\right)$.
Following matrix multiplication rules, operators act from the left
on states in the same row, and from the right on states in the same
column.} \label{states}
\end{figure}

\subsubsection{States}

States contain  scalars, fermions and vectors. Only
the first two are considered in this Section. The matrix space admits two Higgs
doublets $\boldsymbol{\phi}_{1}$ and $\boldsymbol{\phi}_{2}$ (Table 1 and Fig. \ref{states}).
They satisfy $\boldsymbol{\phi}_{1}=\tilde{\gamma}_{5}\boldsymbol{\phi}_{2}$.
Their connection to  Hermitian   and SU(2)  conjugates
is clarified in Section 4.3.

\noindent
\begin{table}[t]
\begin{centering}
\begin{tabular}{cccc}
\toprule
Baryon number zero, Higgs-like scalars  & $I_{3}$  & $Y_o$  & $Q$\tabularnewline
\midrule
\midrule
$\boldsymbol{\phi}_{1}=\begin{pmatrix}\phi_{1}^{+}\\
\phi_{1}^{0}
\end{pmatrix}=\begin{pmatrix}\frac{1}{8}\left(1-i\gamma^{5}\gamma^{6}\right)\left(\gamma^{7}+i\gamma^{8}\right)\gamma_{0}\\
\frac{1}{8}\left(1-i\gamma^{5}\gamma^{6}\right)\left(1+i\gamma^{7}\gamma^{8}\tilde{\gamma}_{5}\right)\gamma_{0}
\end{pmatrix}$  & $\begin{array}{r}
1/2\\
-1/2
\end{array}$  & $1$  & $\begin{array}{r}
1\\
0
\end{array}$\tabularnewline
\midrule
$\boldsymbol{\phi}_{2}=\begin{pmatrix}\phi_{2}^{+}\\
\phi_{2}^{0}
\end{pmatrix}=\begin{pmatrix}\frac{1}{8}\left(1-i\gamma^{5}\gamma^{6}\right)\left(\gamma^{7}+i\gamma^{8}\right)\tilde{\gamma}_{5}\gamma_{0}\\
\frac{i}{8}\left(1-i\gamma^{5}\gamma^{6}\right)\left(1+i\gamma^{7}\gamma^{8}\tilde{\gamma}_{5}\right)\gamma^{7}\gamma^{8}\gamma_{0}
\end{pmatrix}$  & $\begin{array}{r}
1/2\\
-1/2
\end{array}$  & $1$  & $\begin{array}{r}
1\\
0
\end{array}$\tabularnewline
\bottomrule
\end{tabular}
\par\end{centering}
\caption{Scalar Higgs-like doublets} \label{Scalars}
\end{table}

 Non-Higgs scalars can also be constructed  that contribute to the diagonalization
of massive states. Ref. \cite{Romero} provides further information on their nature and their application to obtain fermion properties.

 The massless-fermion  states satisfy the general structure
 of Eq. \ref{eq:10},
and have    massless quark quantum numbers, when classified by
baryon number, isospin, and hypercharge. The matrix space admits four
generations of quarks of different flavor (Fig. \ref{states}), arranged in four
$\textrm{SU}(2)_{L}$ doublets and eight right-handed singlets, shown
in Tables \ref{tab:masslesslefthanded}, \ref{tab:masslessrighthanded}, respectively. After electroweak symmetry breaking, the  Higgs   generates a mass operator used in Section 5 to obtain fermion mass states.

\begin{table}[H]
\begin{centering}
\begin{tabular}{>{\raggedright}p{0.61\textwidth}>{\centering}p{0.06\textwidth}c>{\centering}m{0.06\textwidth}cc}
\toprule
Baryon number 1/3, hypercharge 1/3 and polarization 1/2 $\left(\text{operator }\frac{3}{2}iB\gamma^{1}\gamma^{2}\right)$,
left-handed quark doublets & $I_3$  & $Q$  & $f_{3}$  & $\hat{f}_{3}$  & $F$\tabularnewline
\midrule
\midrule
{\scriptsize{}$\mathbf{Q}_{L1}^{1}=\begin{pmatrix}U_{L1}^{1}\\
D_{L1}^{1}
\end{pmatrix}=\begin{pmatrix}\frac{1}{16}\left(1-\tilde{\gamma}_{5}\right)\left(\gamma^{5}-i\gamma^{6}\right)\left(\gamma^{7}+i\gamma^{8}\right)\left(\gamma^{0}+\gamma^{3}\right)\\
\frac{1}{16}\left(1-\tilde{\gamma}_{5}\right)\left(\gamma^{5}-i\gamma^{6}\right)\left(1-i\gamma^{7}\gamma^{8}\right)\left(\gamma^{0}+\gamma^{3}\right)
\end{pmatrix}$ } & $\begin{array}{r}
1/2\\
-1/2
\end{array}$  & $\begin{array}{r}
2/3\\
-1/3
\end{array}$  & $\begin{array}{r}
1/2\\
1/2
\end{array}$  & $0$ & $\begin{array}{r}
3/2\\
3/2
\end{array}$\tabularnewline
\midrule
{\scriptsize{}$\mathbf{Q}_{L2}^{1}=\begin{pmatrix}U_{L2}^{1}\\
D_{L2}^{1}
\end{pmatrix}=\begin{pmatrix}\frac{1}{16}\left(1-\tilde{\gamma}_{5}\right)\left(\gamma^{5}-i\gamma^{6}\right)\left(1+i\gamma^{7}\gamma^{8}\right)\left(\gamma^{0}+\gamma^{3}\right)\\
\frac{1}{16}\left(1-\tilde{\gamma}_{5}\right)\left(\gamma^{5}-i\gamma^{6}\right)\left(\gamma^{7}-i\gamma^{8}\right)\left(\gamma^{0}+\gamma^{3}\right)
\end{pmatrix}$ } & $\begin{array}{r}
1/2\\
-1/2
\end{array}$  & $\begin{array}{r}
2/3\\
-1/3
\end{array}$  & $\begin{array}{r}
 -1/2\\
-1/2
\end{array}$  & $0$ & $\begin{array}{r}
 -1/2\\
 -1/2
\end{array}$\tabularnewline
\midrule
{\scriptsize{}$\mathbf{Q}_{L3}^{1}=\begin{pmatrix}U_{L3}^{1}\\
D_{L3}^{1}
\end{pmatrix}=\begin{pmatrix}\frac{1}{16}\left(1-\tilde{\gamma}_{5}\right)\left(\gamma^{5}-i\gamma^{6}\right)\left(\gamma^{7}+i\gamma^{8}\right)\gamma^{0}\left(\gamma^{0}-\gamma^{3}\right)\\
\frac{1}{16}\left(1-\tilde{\gamma}_{5}\right)\left(\gamma^{5}-i\gamma^{6}\right)\left(1-i\gamma^{7}\gamma^{8}\right)\gamma^{0}\left(\gamma^{0}-\gamma^{3}\right)
\end{pmatrix}$ } & $\begin{array}{r}
1/2\\
-1/2
\end{array}$  & $\begin{array}{r}
2/3\\
-1/3
\end{array}$  & $\,\,\,\,\,\,0$ & $\begin{array}{r}
1/2\\
1/2
\end{array}$  & $\begin{array}{r}
1\\
1
\end{array}$\tabularnewline
\midrule
{\scriptsize{}$\mathbf{Q}_{L4}^{1}=\begin{pmatrix}U_{L4}^{1}\\
D_{L4}^{1}
\end{pmatrix}=\begin{pmatrix}\frac{1}{16}\left(1-\tilde{\gamma}_{5}\right)\left(\gamma^{5}-i\gamma^{6}\right)\left(1+i\gamma^{7}\gamma^{8}\right)\gamma^{0}\left(\gamma^{0}-\gamma^{3}\right)\\
\frac{1}{16}\left(1-\tilde{\gamma}_{5}\right)\left(\gamma^{5}-i\gamma^{6}\right)\left(\gamma^{7}-i\gamma^{8}\right)\gamma^{0}\left(\gamma^{0}-\gamma^{3}\right)
\end{pmatrix}$}  & $\begin{array}{r}
1/2\\
-1/2
\end{array}$  & $\begin{array}{r}
2/3\\
-1/3
\end{array}$  & $\,\,\,\,\,\,0$  & $\begin{array}{r}
\,\,\, -1/2\\
 -1/2
\end{array}$  & $0$\tabularnewline
\bottomrule
\end{tabular}
\par\end{centering}
\caption{Massless left-handed quark weak isospin doublets. Gauge and Lorentz
operators act from the left and trivially from the right. To obtain the $-1/2$ polarization,
the replacement must be made $\left(\gamma^{0}+\gamma^{3}\right)\rightarrow\left(\gamma^{1}-i\gamma^{2}\right)$,
for $\mathbf{Q}_{L1}^{1}$, $\mathbf{Q}_{L2}^{1}$, and $\left(\gamma^{0}-\gamma^{3}\right)\rightarrow\left(\gamma^{1}-i\gamma^{2}\right)$,
for $\mathbf{Q}_{L3}^{1}$, $\mathbf{Q}_{L4}^{1}$. }
\label{tab:masslesslefthanded}
\end{table}

\begin{table}[H]
\begin{centering}
\begin{tabular}{>{\raggedright}p{0.6\textwidth}>{\centering}p{0.06\textwidth}c>{\centering}p{0.06\textwidth}cc}
\toprule
Baryon number 1/3 and polarization 1/2

$\left(\text{operator }\frac{3}{2}iB\gamma^{1}\gamma^{2}\right)$,
right-handed quark singlets& $Y_o$  & $Q$  & $f_{3}$  & $\hat{f}_{3}$  & $F$\tabularnewline
\midrule
\midrule
$\begin{array}{l}
U_{R1}^{1}=\frac{1}{16}\left(1+\tilde{\gamma}_{5}\right)\left(\gamma^{5}-i\gamma^{6}\right)\left(\gamma^{7}+i\gamma^{8}\right)\gamma^{0}\left(\gamma^{0}+\gamma^{3}\right)\\
D_{R1}^{1}=\frac{1}{16}\left(1+\tilde{\gamma}_{5}\right)\left(\gamma^{5}-i\gamma^{6}\right)\left(1-i\gamma^{7}\gamma^{8}\right)\gamma^{0}\left(\gamma^{0}+\gamma^{3}\right)
\end{array}$  & $\begin{array}{r}
4/3\\
-2/3
\end{array}$  & $\begin{array}{r}
2/3\\
-1/3
\end{array}$  & $\begin{array}{r}
 1/2\\
 1/2
\end{array}$  & $0$  & $\begin{array}{r}
3/2\\
3/2
\end{array}$\tabularnewline
\midrule
$\begin{array}{l}
U_{R2}^{1}=\frac{1}{16}\left(1+\tilde{\gamma}_{5}\right)\left(\gamma^{5}-i\gamma^{6}\right)\left(1+i\gamma^{7}\gamma^{8}\right)\gamma^{0}\left(\gamma^{0}+\gamma^{3}\right)\\
D_{R2}^{1}=\frac{1}{16}\left(1+\tilde{\gamma}_{5}\right)\left(\gamma^{5}-i\gamma^{6}\right)\left(\gamma^{7}-i\gamma^{8}\right)\gamma^{0}\left(\gamma^{0}+\gamma^{3}\right)
\end{array}$  & $\begin{array}{r}
4/3\\
-2/3
\end{array}$  & $\begin{array}{r}
2/3\\
-1/3
\end{array}$  & $\begin{array}{r}
-1/2\\
-1/2
\end{array}$  & $0$  & $\begin{array}{r}
-1/2\\
-1/2
\end{array}$\tabularnewline
\midrule
$\begin{array}{l}
U_{R3}^{1}=\frac{1}{16}\left(1+\tilde{\gamma}_{5}\right)\left(\gamma^{5}-i\gamma^{6}\right)\left(\gamma^{7}+i\gamma^{8}\right)\left(\gamma^{0}-\gamma^{3}\right)\\
D_{R3}^{1}=\frac{1}{16}\left(1+\tilde{\gamma}_{5}\right)\left(\gamma^{5}-i\gamma^{6}\right)\left(1-i\gamma^{7}\gamma^{8}\right)\left(\gamma^{0}-\gamma^{3}\right)
\end{array}$  & $\begin{array}{r}
4/3\\
-2/3
\end{array}$  & $\begin{array}{r}
2/3\\
-1/3
\end{array}$  & $0$  & $\begin{array}{r}
1/2\\
1/2
\end{array}$  & $\begin{array}{r}
1\\
1
\end{array}$\tabularnewline
\midrule
$\begin{array}{l}
U_{R4}^{1}=\frac{1}{16}\left(1+\tilde{\gamma}_{5}\right)\left(\gamma^{5}-i\gamma^{6}\right)\left(1+i\gamma^{7}\gamma^{8}\right)\left(\gamma^{0}-\gamma^{3}\right)\\
D_{R4}^{1}=\frac{1}{16}\left(1+\tilde{\gamma}_{5}\right)\left(\gamma^{5}-i\gamma^{6}\right)\left(\gamma^{7}-i\gamma^{8}\right)\left(\gamma^{0}-\gamma^{3}\right)
\end{array}$  & $\begin{array}{r}
4/3\\
-2/3
\end{array}$  & $\begin{array}{r}
2/3\\
-1/3
\end{array}$  & $0$  & $\begin{array}{r}
-1/2\\
-1/2
\end{array}$  & $0$\tabularnewline
\bottomrule
\end{tabular}
\par\end{centering}
\centering{}\caption{Massless right-handed quark weak isospin singlets. Gauge and Lorentz
operators act from the left and trivially from the right. To obtain the $-1/2$ polarization,
the replacement must be made $\left(\gamma^{0}+\gamma^{3}\right)\rightarrow\left(\gamma^{1}-i\gamma^{2}\right)$,
for $U_{R1}^{1}$, $U_{R2}^{1}$, $D_{R1}^{1}$, $D_{R2}^{1}$,
and $\left(\gamma^{0}-\gamma^{3}\right)\rightarrow\left(\gamma^{1}-i\gamma^{2}\right)$,
for $U_{R3}^{1}$, $U_{R4}^{1}$, $D_{R3}^{1}$, $D_{R4}^{1}$.}
\label{tab:masslessrighthanded}
\end{table}

\subsection{Fermion Yukawa elements}

Bilinear fermion terms    can  be constructed that  produce scalar elements
transforming quarks into their different combinations. We use the   (7+1)-d space represented in Fig. \ref{states},  with     particular and general  properties that  can be distinguished.

There are two matrix configurations:
\begin{equation}
 \label{YukawaItz}
 P^{ F\alpha\beta}_{i}={Q_{Ri}^{\alpha}} {\bar Q}_{Li}^{\beta}   \ \ \ \ \  i=1,2,3\ \ \
\end{equation}
  is contained in the Dirac projector with  ($\alpha$, $\beta$)-spin components and (positive or negative)-energy; the  three $P^{ F\alpha\beta}_{i}$  are the same up to a phase;
   $Q$ are   $U$- or $D$-type fermions  obtained from Tables \ref{tab:masslesslefthanded},   \ref{tab:masslessrighthanded},  defining $F$, the  $R$, $L$ case taken as an example,   and   ${{\bar Q}_{Li}^{\alpha}={ Q}_{Li}^{\alpha}}^\dagger\gamma_0^B$;
\begin{eqnarray}
\label  {gammaB0}
    { \gamma_0^B}    = 2( \phi_1^0 +{\phi_1^0}^\dagger) ,
 \end{eqnarray}
 $\phi_1^0$ defined in Table \ref{Scalars}.
The $i$, $j$  imply we choose a 3-generation (arbitrary) projection
 to reproduce the SM; we also note that   ${{ Q}_{Ri}^{\alpha}}  {{\bar Q}_{Lj}}^{\beta} =0$ for $i \neq j$.

On the other hand,
\begin{equation}
 \label{YukawaOp}
 Y^{ F}_{ij}= {{\bar Q}_{Ri}}^{\alpha}  {Q_{Lj}^{\alpha}} \ \ \ \ \  i,j=1,2,3
\end{equation}
   defines the   Yukawa basis (full flavor transition matrix) to be used in Section 6, for the complete scalar-fermion  SM Lagrangian component.
  One can check that   $Y^{ U}_{ij}$, $Y^{ D}_{ij}$  are the same  (up to phases), so they are commonly labelled   $Y^{ F}_{ij}$.
   The set $R$, $L$, $\alpha$  is arbitrary and other choices will reproduce (up to phases) the nine  $Y^{ F}_{ij}$ terms.
   Indeed, although the (7+1)-d basis can accommodate  four generations, the projection operator for, say, flavors 1,2,3
   \begin{equation}
 \label{ProjeFla}
  Y^{F4}=  Y^{ F}_{11}+ Y^{ F}_{22} + Y^{ F}_{33},
    \end{equation}
 induces  the 3-generation subset with 9 elements, ${Y^{ F4}}^\dagger Y^{ F}_{ij} {Y^{F4}}$ .  As the set is closed under matrix multiplication,
 the 4th generation is discarded (see Section 6.)



 The resulting  projection operators may be understood from the products of a fermion with  matrix  structure    $|{\rm spin}\rangle  \langle {\rm flavor} |$  and an hermitian conjugate one,
 resulting in the form $|{\rm spin}\rangle  \langle {\rm spin} |$ for  Eq. \ref{YukawaItz}, and, inverting the order,  $|{\rm flavor}\rangle  \langle {\rm flavor} |$ in Eq. \ref{YukawaOp}.

\section{Fermion-vector Lagrangian: chiral basis in spin space}

Concentrating on the heaviest fermions, the SM two-quark\footnote{A single generation is used, and CKM mixing is neglected; Eq. \ref{Bilin} describes the electroweak interaction for   one quark color, and a sum is assumed over each  such term. }  electroweak  interaction  Lagrangian\cite{Glashow} is\footnote{We use units with $\hbar=c=1$, and metric $g_{\mu\nu}=(1,-1,-1,-1) $ throughout.}
\begin{eqnarray}
\label {Bilin}
  {\cal L}_{FV}=  {\bar {  \bf q }_L}(x) [ i\partial_\mu + \frac{1}{2}g\tau^a  W_\mu^a(x)  +\frac{1}{6}g'  B_\mu (x) ] \gamma^\mu  { \bf q}_L (x)     + \nonumber\\{ \bar   t_R      }(x)      [ i\partial_\mu   +\frac{2}{3}g'  B_\mu (x) ]    \gamma^\mu    t_{R}(x)
 +{ \bar   b_R  (x)    }      [ i\partial_\mu    -\frac{1}{3}g'  B_\mu (x) ]    \gamma^\mu    b_{R}(x),
\end{eqnarray}
where   the   spin-1/2  fields  consist of $  {\bf q}_{L}(x)=\left(\begin{array}{lcr}
   t_{L}(x)  \\
  b_{L}(x)
 \end{array}\right)$, a  left-handed        hypercharge $Y =1/3$  SU(2)$_L$-doublet,  and  $t_{R }(x)$, $b_{R }(x)$,  right-handed  $Y  = 4/3, -2/3$ singlets, respectively; each  term contains two  polarizations   as,  e. g., $     t_{L}(x)=\left(\begin{array}{lcr}
     \psi_{tL}^1(x) \\
  \psi_{tL}^2(x)
 \end{array}\right)$;    $\psi_{qh}^\alpha(x)$ are wave functions\footnote{For simplicity, spin and scalar representations are assumed that give the states' form.}  for quarks $q=t,b,$ with spin components $\alpha=1,2$, and chirality  $h=L,R$;   $W_\mu^a(x)$, $a=1,2,3$, and  $B_\mu (x)$,  are    associated gauge-group weak and hypercharge vector bosons, with  coupling constants,  $g$, $g'$,
 respectively;  $\tau^a  $  are the Pauli matrices representing the  SU(2)$_L$  generators.

 An extended (7+1)-d   Clifford algebra comprises a sufficiently large space to describe heavy SM particles\cite{BesproRicardo,Romero}, with the 4-d Lorentz symmetry maintained, and  spin-component   generators     $\frac{3}{2}B\sigma_{\mu\nu}$,  where
$ \label {sigmamunu}
 \sigma_{\mu\nu}=\frac{i}{2}[\gamma_\mu ,\gamma_\nu ],
 $
   and $\mu ,\nu =0,...,3;$ additional scalar generators use $\gamma_{5}, ..., \gamma_{8}$,   producing the
baryon-number   operator $B$ in Eq. \ref{eq:2.5.4},
  which   conforms a  spin-space projection partition,  and   gives quarks
   1/3 ($-1/3$ for  antiparticles,) and  bosons 0.

Other scalar-symmetry  generators
 include   the hypercharge
$  Y_o$ in Eq.   \ref{eq:2.5.5},
 with $\tilde{\gamma}_{5}=-i\gamma_{0}\gamma_{1}\gamma_{2}\gamma_{3}$,
the weak SU(2)$_L$ terms in Eq.  \ref{eq:2.5.6}
  and  flavor generators in Eqs.  \ref{eq:2.5.8}-\ref{eq:2.5.11};
  as required, $[I_i,I_j]=i\epsilon_{ijk}I_k$, $[I_i,Y_o]=[B, Y_o]=[B,I_i]=[3 B \sigma_{\mu\nu},Y_o]=[3 B \sigma_{\mu\nu},I_i]=0$.

The (7+1)-d  space  allows    for   a description of  quark fields
\begin{eqnarray} \label{quarkfields}\boldmathPsi_{qL}(x)=\sum_\alpha
   \psi_{tL}^\alpha(x) T_{L}^\alpha  +
 \psi_{bL}^\alpha(x) B_{L}^\alpha,    \\ \nonumber
  { \Psi_{tR} }(x)=\sum_\alpha \psi_{tR}^\alpha(x)T^\alpha_R, \  { \Psi_{bR}  }(x)=\sum_\alpha \psi_{bR}^\alpha (x)B^\alpha_R,
  \end{eqnarray} with hypercharges $1/3$,  $4/3$,      $-2/3$, respectively,
   and spinor components chosen in Table \ref{tab:tableqbquarks}, given explicitly in Tables \ref {tab:masslesslefthanded}, \ref {tab:masslessrighthanded}; the quantum numbers $\lambda$ are obtained from
 the operator structure $[Op,\Psi]=\lambda  \Psi   $ for the weak component $I_3$,  hypercharge $Y_o$ (or charge $Q=I_3+\frac{1}{2}Y_o$,) and   spin-polarization $\frac{3i}{2}B\gamma^{1}\gamma^{2}$\ operators.

The  SM     Lagrangian ${\cal L}_{FV}$ in Eq. \ref{Bilin} can be equivalently written\footnote{The commutator is omitted as the operator acts trivially on one side.}  in this basis:
as derived in Ref. \cite{BesproRicardo}, and examined in Ref. \cite{BesproRicardoNuc}
\begin{eqnarray}
\label {BilinNoTracElectroweak}
   {\cal L}_{FV}=
  {\rm tr } \{ { \boldmathPsi_{qL}   ^\dagger     }(x)       [ i\partial_\mu +  g I^a  W_\mu^a(x)  + \frac{1}{2} g'   Y_o  B_\mu (x) ]  \gamma^0  \gamma^\mu   \boldmathPsi_{qL} (x) + \nonumber \\
         \Psi _{tR } ^\dagger (x)       [ i\partial_\mu   +  \frac{1}{2}  g' Y_o B_\mu (x) ]    \gamma^0 \gamma^\mu     {     \Psi _{tR }}(x)
 +{     \Psi}_{bR}  ^\dagger(x)             [ i\partial_\mu   +\frac{1}{2} g'  Y_o  B_\mu (x) ]   \gamma^0 \gamma^\mu    { \Psi_{bR} }(x)  \}P_f ,
\end{eqnarray}
while gauge and Lorentz symmetries can be checked with the above transformation rule, or given  the equivalence to the traditional formulation. A projection operator $P_f$  that connects the two expressions\cite{BesproRicardoNuc} can be omitted by finding phases for $\boldmathPsi$, which translates into finding an adequate $\gamma_\mu$ basis. The trace coefficient is usually 1, as the field normalization factor accounts for reducible representations. A complete proof of the equivalence is given in Appendix 1.
\begin{table}[ht]
\begin{centering}
\begin{tabular}{|c|c|c|c|}
\hline
(a) hypercharge  $1/3$   left-handed doublet & $I_{3}$ & $Q$ & $\frac{3i}{2}B\gamma^{1}\gamma^{2}$\tabularnewline
\hline
\hline
$ \left(\begin{array}{lcr}
   T_{L}^1 \\
  B_{L}^1\\
 \end{array}\right)=\left(\begin{array}{c}
 U_{L1}^1
 \\
  D_{L1}^1
\end{array}\right)$ & $\begin{array}{r}
1/2\\
-1/2
\end{array}$ & $\begin{array}{r}
2/3\\
-1/3
\end{array}$ & $\begin{array}{r}
 1/2\\
 1/2
\end{array}$\tabularnewline
\hline
\end{tabular}
\center{(a)}
\par\end{centering}
\begin{centering}
\begin{tabular}{|c|c|c|c|}
\hline
(b)   $I_{3}=0$   right-handed singlets & $Y$ & $Q$ & $\frac{3i}{2}B\gamma^{1}\gamma^{2}$\tabularnewline
\hline
\hline
\ \ \ \  \ \ \ \ $\begin{array}{c}
T_R^1=U_{R1}^1 \\
B_R^1 =
 D_{R1}^1
\end{array}$ & $\begin{array}{r}
4/3\\
-2/3
\end{array}$ & $\begin{array}{r}
2/3\\
-1/3
\end{array}$ & $\begin{array}{r}
 1/2\\
 1/2
\end{array}$\tabularnewline
\hline
\end{tabular}
\end{centering}
\center {(b)}
 \caption {(a) Quantum numbers of massless  left-handed quark weak isospin doublet, and (b) right-handed  singlets, with momentum along $\pm{\bf {\hat z}}$,
given explicitly in Tables \ref{tab:masslesslefthanded}, \ref {tab:masslessrighthanded}. The spin component along $\hat {\bf z}$,
$
i\frac{3}{2}B\gamma^{1}\gamma^{2},
$ is used.}
\label{tab:tableqbquarks}
\end{table}




The
   W-fermion   vertex in ${\cal L}_{FV}$, Eq. \ref{BilinNoTracElectroweak},
    contains the matrix element    $ \langle F' |W_{o\mu}^i | F   \rangle$,    where    the
W contribution    \begin{eqnarray} \label{Wdef}
  W_{o\mu} ^i   =g \gamma_0 \gamma_\mu I ^i
 \end{eqnarray}    describes the SU(2)$_L$ inherently  chiral  action  on   fermion states $ | F   \rangle, | F'   \rangle$,
   as  it carries   the    projection $L_5=\frac{1}{2}(1-\tilde\gamma_5)$, predicted by the spin basis\cite{Romero};
  it is thus the    natural choice. For example,
   this property is absent for  $W _{o\mu} ^{\prime\ i}   =g \gamma_0 \gamma_\mu I ^{\prime}_i$, where $I ^{\prime}_i $ are the SU(2)$_L$ generators without  $L_5$;  although  an equivalent interaction term    results
   within this space,    it requires the inclusion of  $L_5$   within the vertex; worse,   $ [Y_{o},I^{\prime}_ {1,2}] \neq 0$.

\section{Scalar-vector Lagrangian: extended charge-conjugate symmetry }
\subsection{Conventional $ {\cal L}_{SV}$ }
In the  SM, the  Higgs particle is present\cite{Glashow} in the  SU(2)$_L\times $U(1)$_Y$  gauge-invariant interacting Lagrangian-density component
\begin{eqnarray}\label{LSVConventinal}  {\cal L}_{SV}= {\bf H}^\dagger(x) {{\bf F}^\mu}^\dagger (x)  {\bf F}_\mu(x) {\bf H}(x),\end{eqnarray} with \begin{eqnarray}\label{F}{\bf F}_\mu(x)= i \partial_\mu+\frac{1}{2}g  {\boldmathtau}\cdot {\bf W_\mu}(x)+\frac{1}{2}g'B_\mu(x),\end{eqnarray} 
 ${\bf W_\mu}(x)=(W^1_\mu(x),W^2_\mu(x),W^3_\mu(x))$,   and  the $Y=1$
  complex-doublet  scalar
  ${\bf H}(x)=\frac{1}{\sqrt{2}} \left(\begin{array}{c}
  \eta_1(x)+i\eta_2(x) \\
  \eta_3(x)+i\eta_4(x)
\end{array}\right)$, composed of two  charged (upper), and two neutral  (lower)  fields.
\subsection{${\cal L}_{SV}$ with Higgs and conjugate }
 ${\cal L}_{SV}$ can be equivalently written   (with $B_\mu(x)\rightarrow -B_\mu(x)$)  in terms of  the orthogonal  $Y=-1$ combination   $\tilde{\bf H}(x)=i \tau_2 {\bf H}^*(x)$,  which uses an antiunitary  transformation ${\cal C}$  expressing charge-conjugation invariance   (in addition to the  CP symmetry in the electroweak sector, and approximate SU(2)$_L\times $SU(2)$_R$ symmetry\cite{Weinstein}; a Hilbert space is assumed;)    this is also a consequence of the    SU(2)  property that the conjugate representation is obtained from a similarity transformation, which
  ensures independence of the doublet choice.
   Appendix 2 shows that
 \begin{eqnarray}\label{LSVHiggsAndScalar} {\cal L}_{SV}= {\rm tr}{[ {\bf F'}_\mu  \bar{\bf H}_{\chi_t   \chi_b}(x)}]^\dagger {\bf F'}^\mu \bar{\bf H}_{\chi_t   \chi_b}(x),  \end{eqnarray}  where
  $\bar{\bf H}_{ \chi_t    \chi_b }(x)=(\chi_t  {\bf H}(x),\chi_b \tilde{\bf H}(x))$  is a $4 \times 4$
matrix, $\chi_t$, $\chi_b$ are
 complex, and $|\chi_t|^2+|\chi_b|^2=1$,
  with
  \begin{eqnarray}\label{Fp}{\bf F'}_\mu \bar{\bf H}_{\chi_t   \chi_b }(x) =
 (i \partial_\mu+\frac{1}{2}g  {\boldmathtau}\cdot {\bf W_\mu}(x)) \bar{\bf H}_{\chi_t   \chi_b }(x) +
g'\bar{\bf H}_{\chi_t   \chi_b }(x) B_\mu(x)\tau_3,\end{eqnarray}
  which is diagonal in   $ {\bf H}(x)$,  $\tilde{\bf H}(x)$, and hence does not mix them. Moreover ${\cal L}_{SV}$ is a sum of weighted positive-definite terms, meaning only the combination $|\chi_t|^2+|\chi_b|^2$ results.   This generalizes the expression\cite{Sikivie,Chivukula}  for ${\cal L}_{SV}$ in terms of
  $\bar {\bf H}_{ \frac{1}{\sqrt{2}} \frac{1}{\sqrt{2}}}(x) $.  With the U(1) overall phase, a three-parameter subspace of the norm-conserving constraint $|\chi_t|^2+|\chi_b|^2=1$ is generated. We associate this isometry with the   ${\cal L}_{SV}$  invariance under ${\cal C}$:  $ -\tau_2{\cal K} {\bf F'}_\mu \bar{\bf H}_{\chi_t   \chi_b }(x) \tau_2{\cal K}=
  {\bf F'}_\mu \bar{\bf H}_{\chi_b^*\chi_t^*}(x)$, with ${\cal K}$ the complex conjugate operator;
     ${\cal L}_{SV}$ is also invariant under the  ${\cal \tau}_3$  transformation defined as
   $\bar{\bf H}_{\chi_t   \chi_b}(x) \rightarrow \bar{\bf H}_{\chi_t   \chi_b}(x) \tau_3 $,   together with the combination     ${\cal C}\tau_3$.

  Further extension   can   be made for the scalars in the spin  basis by attaching   the $\tilde \gamma_5$ operator. Using the   projection operators in Eq. \ref {projeRL},
  ${\cal L}_{SV}$ in  Eq.  \ref{LSVHiggsAndScalar}  is   generalized with the substitutions
    \begin{eqnarray} \label{substitu}
   {\bf F'}_\mu\rightarrow (L_ 5)_ {4\times4} {\bf F'}_\mu\\
  \bar{\bf H} \rightarrow (L_ 5) _{4\times4} ( \gamma_ 0)_ {4\times4}  \bar{\bf H} ,
    \end{eqnarray}
  thus  including spin degrees of freedom, leading to   a combined spinor-electroweak description.
 An intermediate expression that connects to the spin basis, and ultimately to   Yukawa  components,  is obtained
\begin{eqnarray}\label{LSVHiggschi} {\cal L}_{SV}= \frac{1}{2}{\rm tr}{[ L_5 {\bf F'}_\mu  L_5\gamma_0 \bar{\bf H}_{\chi_t   \chi_b}(x)}]^\dagger L_5 {\bf F'}^\mu   L_5\gamma_0 \bar{\bf H}_{\chi_t   \chi_b}(x)\\
=\frac{1}{4}{{\rm tr} ([ L_5 {\bf F'}_\mu  L_5\gamma_0 \bar{\bf H}_{\chi_t   \chi_b}(x)}]^\dagger+ L_5 {\bf F'}_\mu   L_5\gamma_0 \bar{\bf H}_{\chi_t   \chi_b}(x)) \\  \nonumber
( L_5 {\bf F'}^\mu   L_5\gamma_0 \bar{\bf H}_{\chi_t   \chi_b}(x)+{[ L_5 {\bf F'}^\mu  L_5\gamma_0 \bar{\bf H}_{\chi_t   \chi_b}(x)}]^\dagger),  \end{eqnarray}
with the trace also over spin degrees of freedom,   the second equality using  hermitian conjugates, $ R_5L_5=L_5 R_5=0$, and trace properties which lead  to only two identical non-trivial terms.
These forms will prove
  useful in comparing with Yukawa terms below.

 \subsection{${\cal L}_{SV}$ in (7+1)-d spin space}
 In the spin basis,   the four-scalar doublet structure above is reproduced. Indeed,    it
emerges naturally in the (7+1)-d spin basis,   with the Higgs potential not altered under different definitions (chiral ones or not.)
    Table \ref{Scalars}    presents    two  of these scalar elements (with two additional as their conjugates.)  Together with coordinate dependence, they are

    \begin{eqnarray} \label{phibolddef}
 \boldsymbol{\phi}_1(x)& = & \frac{1}{\sqrt{2}} %
\left [ \eta_1(x)+i\eta_2(x)   \right ] \phi_{1}^{+} +
   \frac{1}{\sqrt{2}}  \left  [ \eta_3(x)+i\eta_4(x) \right ]   \phi_{1}^{0} \nonumber  \\
 { \boldmathphi}_2(x)
&=& \frac{1}{\sqrt{2}}
 \left [ \eta_1(x)+i\eta_2(x)   \right ] \phi_{2}^{+}
 + \frac{1}{\sqrt{2}}    \left  [ \eta_3(x)+i\eta_4(x) \right ]   \phi_{2}^{0}
 ,  \end{eqnarray}
  and   whose quantum numbers associate them to the Higgs doublet.  These are unique within the  (7+1)-d space\cite{Romero}. Although   new scalar fields are introduced in principle, here we concentrate on the SM-equivalent projections.
Given the SM Higgs    conjugate representation
$\tilde{\bf H}(x)$
 the   scalar components are interpreted    through the assignments (see Table \ref{Scalars}),
\begin{eqnarray}
\label {correspH}
 { \bf H(x)} \rightarrow \boldmathphi_1(x)-\boldmathphi_2(x) \nonumber \\
 \tilde{\bf H}^\dagger(x) \rightarrow\boldmathphi_1(x)+\boldmathphi_2(x).
\end{eqnarray}
This leads to the equivalent expressions
\begin{eqnarray}\label{EquivSV}
   {\cal L}_{SV}&=& { \rm tr  }\{[{\bf F''} (x) ,{\bf H}_{af}(x) ]_\pm^\dagger
[ {\bf F''} (x) ,{\bf H}_{af}(x)  ]_\pm \}_{\rm sym} \\
&=&  \frac{1}{2}  { \rm tr  }\{[{\bf F''}(x) ,{\bf H}_{af}(x)+ {\bf H}_{af}^\dagger(x) ]_\pm^\dagger
[ {\bf F''}(x) ,{\bf H}_{af}(x)+ {\bf H}_{af}^\dagger(x)  ]_\pm \}_{\rm sym}, \nonumber \\ \label{EquiSVHer}
\end{eqnarray}   where we
introduced ${\bf H}_{af}(x)=a\boldmathphi_1(x)+f \boldmathphi_2(x)$,  and  \begin{eqnarray}\label{Fpp}{\bf F''}(x)=
[i\partial_\mu+    {  g W_\mu^i(x) I_i} +\frac{1}{2}g' B_\mu(x) Y_o]\gamma_0 \gamma^\mu ; \end{eqnarray}   the subindex {\it sym} means only symmetric $\gamma_\mu \gamma_\nu$  components are taken, to avoid the Pauli components; and the  $\pm$
  index means the commutator and  the anticommutator should be used for the  temporal and spatial  $\gamma_\mu$ components,  respectively.
The equality for ${\cal L}_{SV}$  implies   that    it        accommodates SM  parity-conserving scalar representations.
  The complex parameters  $a$, $f$, are constrained by the normalization rule $|a|^2+|f|^2=1$. These properties for ${\cal L}_{SV}$ are shown explicitly in Appendix 2.

\subsection{${\cal L}_{SV}$ mass components in conventional and  (7+1)-d spin space}

The spin  representation can be connected with that of  $\bar{\bf H}_{\chi_t   \chi_b}(x)$ with the expression  ${\bf H}_{af}(x) = \frac{1} {\sqrt{2}}(\chi_t  {\bf H}_t(x)+ \chi_b  {\bf H}_b(x)  )$,   where  \begin{eqnarray}
\label{DefHtHb} {\bf H}_{t}(x)=  { \boldmathphi}_1(x)+{ \boldmathphi}_2(x) \\ \nonumber  {\bf H}_{b}(x)=   { \boldmathphi}_1(x)-{ \boldmathphi}_2(x),  \end{eqnarray}  with ${ \boldmathphi}_i $   defined in   Eq.   \ref{phibolddef}, and  this  parameterization applies the  $\it unitary$ transformation
    $\chi_t=\frac{1} {\sqrt{2}}(a+f)$, $\chi_b=\frac{1} {\sqrt{2}}(a-f)$.

 Under the Higgs mechanism,   the  SM
scalars acquire\cite{HiggsMech,HiggsMechHiggs}       a  vacuum expectation value $v$, and only the neutral field $\eta_3(x)$ survives:  $\langle \eta_3(x)\rangle=v$,
 $\langle {\bf H}(x)\rangle=  \frac{v}{\sqrt{2}} \left(\begin{array}{c}
 0\\
  1
\end{array}\right)$,
 while the charged and imaginary components are absorbed into vector bosons,
as seen explicitly in the unitary gauge.  Idem in the spin basis, as can be  proved by the Lagrangian equivalence or directly; then,
 \begin{eqnarray}
\label  {HiggsComponentsYm1}
  \langle{\bf H}_{af}(x)\rangle= H_n= \frac{v} {2} (\chi_t  H_t^0+\chi_b  H_b^0  ),
 \end{eqnarray}
 where the normalized Higgs operator $H_n$ is defined,   with the same  $0,+$   component  conventions as  for  the ${ \boldmathphi}_i $,
  implying,  as  ${\rm tr}{ H^0}_{i}^\dagger{ H^0}_{j}=2 \delta_{ij}$, $i,j=t,b$,
 \begin{eqnarray}
\label  {HiggsNormVacuumExpRR}
  \langle {\bf H}_{af}^\dagger(x){\bf H}_{af}(x) \rangle =(|a|^2+|f|^2)v^2/2   =(|\chi_t|^2+|\chi_b|^2)v^2/2=v^2/2.
  \end{eqnarray}


The vector-Higgs vertex in  ${\cal L}_{SV}$ determines  the vector-boson masses, and within the  spin basis,
the trace     is taken consistently with
     $H_{n}$.
  Thus, the    mass component, extracted from Eq. \ref{EquivSV}, taking for ${\bf F''}$ the W, Z field terms, and for ${\bf H}_{ab}$ its vacuum expectation value in Eq.  \ref{HiggsComponentsYm1},
 \begin{eqnarray}
\label  {HiggsCompleteYm1}{\cal L}_{SVm}=  {\rm  tr }([ H_n ,  g W_0^m(x)  I_m+\frac{g'}{2} Y_o  B_0(x)  ]^\dagger
   [ H_n , gW_0^l(x)  I_l+
   \frac{g'}{2} Y_o  B_0(x) ]+ \\ \nonumber
   \{H_n,(g W_i^k(x)  I_k+ \frac{g'}{2}Y_o  B_i(x)) \gamma_0\gamma^i \}^\dagger
  \{ H_n,(g W_j^l(x)  I_l+ \frac{g'}{2} Y_o B_j(x)) \gamma_0\gamma^j \} )
  \end{eqnarray} is produced.
   For the neutral massive vector boson, one derives the normalized  $Z_\mu(x)=(-g W_\mu^3(x)+g' B_\mu(x))/\sqrt{g^2+ {g'}^2}$, and   massless photon $A_\mu(x)=(g' W_\mu^3(x)+g B_\mu(x))/\sqrt{g^2+ {g'}^2}$,
 giving,  e. g., the 0-component
  \begin{eqnarray}
\label {ZmassEq}
  {\cal L}_{SZm0}&= & {\rm tr}[H_n , W_0^3(x) g I_3+ B_0(x) \frac{1}{2}g'Y_o  ]^\dagger[H_n ,  W_0^3(x) g I_3+  B_0(x) \frac{1}{2} g' Y_o ] \\ \nonumber &=&
 Z_0^2(x)\frac{1}{  g^2+ {g}^{\prime 2}  } {\rm tr}[H_n ,   g^2 I_3-   \frac{1}{2}g^{\prime 2}Y_o  ]^\dagger[H_n ,   g^2 I_3-   \frac{1}{2}g^{\prime 2}Y_o]  =\frac{1}{2} Z_0^2(x)    m_Z ^2
 ,
\end{eqnarray}
 implying \begin{eqnarray}
\label {ZmassEqEx}{\rm tr}\frac{1}{  g^2+ {g}^{\prime 2}  }[\sqrt{2}H_n ,   g^2 I_3-   \frac{1}{2}g^{\prime 2}Y_o  ]^\dagger[\sqrt{2}H_n ,   g^2 I_3-   \frac{1}{2}g^{\prime 2}Y_o]  =   v^2 (g^2+ {g'}^2)/4,\end{eqnarray}
thus, $m_Z=v\sqrt{g^2+ {g'}^2}/2,$ $m_A=0.$

Similarly,   for ${\cal L}_{SWm}$, the  $W_{o\mu} ^i$ basis   in Eq. \ref{Wdef}
 emerges, and
 defines the masses of the charged boson fields
   $W_\mu^\pm(x)=\frac{1}{\sqrt{2}}(W_\mu^1(x)\mp i W_\mu^2(x))$. Thus,
the    charged-vector boson component  \begin{eqnarray}
\label {WmassEq}{\cal L}_{SWm0}= W_0^{ i}(x)  W_{0}^{j}(x) {\rm tr}[ H_n, W_{o0}^i]^\dagger [H_n, W_{o0}^j] = m_W^2 W^+_0(x)W^-_0(x) , \end{eqnarray}
$i,j=1,2$ contains $ m_W^2={\rm tr}[H_n, W_{o0}^+]^\dagger [H_n, W_{o0}^+]=v^2g^2/4$, with
 $W^\pm _{o\mu}  =\frac{1}{\sqrt{2}}g\gamma_0\gamma_\mu I^\pm$,  $I^\pm=I_1 \pm iI_2$. This assignment is unique as this is the only way to maintain not only the  vertex condition (gauge invariance,)
 but also  normalization (above.)
When written   in terms of $H=H_n + H_n^\dagger $,   interpreted      as a fermion Hamiltonian,
  $ m_W^2={\rm tr}[H , W_{o0}^+]^\dagger[H , W_{o0}^+]$
and the other part is not affected,
 as $[  H_n^\dagger ,W_{o\mu}^+ ]=0$,


\section{Scalar-fermion Lagrangian: heavy-quark doublet's mass constraint }

The Yukawa fermion-scalar interaction can be similarly parameterized in the Clifford basis
\begin{eqnarray}
\label {BilinHiggsFer}
-{\cal L}_{SF}&=& {\rm tr}
   \frac{\sqrt{2}}{v}  [ m_t   \Psi_{tR} ^\dagger(x) {\bf   H}_{t}(x) { \boldmathPsi}_{qL}(x) +
 m_b {  \boldmathPsi}_{qL}  ^\dagger(x) {\bf  H}_{b}(x) \Psi_{bR} (x)]    +\{ hc \},
\end{eqnarray}
where  $m_t$ and  $m_b$ are the top and bottom masses, respectively, and the fermion fields $\Psi$ are defined in Eq. \ref{quarkfields}.
We note  that the Higgs scalar components  have the  correct chiral  action  over   fermions: under the    projection operators in Eq. \ref{projeRL} $L_5$, and  $R_5$,
 e. g., $R_5{\bf H}_{ t}(x)L_5={\bf H}_{ t}(x)$, $L_5{\bf H}_{ b}(x)R_5={\bf H}_{ b}(x)$, $L_5{\bf H}_{ t}(x)R_5=0$,
$R_5{\bf H}_{ b}(x)L_5=0$.
For Eq. \ref{BilinHiggsFer}, the underlying    mass operator
is ${\bf H}_m(x)=\frac{\sqrt{2}}{v} (m_t {\bf  H}_{t}(x)+m_b {\bf  H}_{b}(x))$, giving,  under the Higgs mechanism, \begin{eqnarray} \label{HmVacuum}  \langle{\bf H}_m(x)  \rangle=H_m =  m_t {  H}_{t}^0 +m_b    H _{b}^0. \end{eqnarray}
Examples of quark massive basis states
 are summarized on Table \ref{Tmassivequark} (see Tables  \ref{tab:masslesslefthanded}-\ref{tab:tableqbquarks}), for both u and d-type quarks, with their quantum numbers.
 Only one polarization and one flavor are shown, as a
more thorough treatment of the fermion-flavor states  are given elsewhere\cite{Romero}.

This results in,
e. g.,
\begin{eqnarray}
\nonumber
 H_m^h T_{M}^1=m_t T_{M}^1, \ \   H_m^h  T_{M}^{c1} =-m_t T_{M}^{c1}  , \\
   \label{MassiveEigenvalues}  H_m^h B_{M}^1 =m_b B_{M}^1 , \  \  H_m^h  B_{M}^{c1} =-m_b B_{M}^{c1},
 \end{eqnarray}
 where $H_m^h=H_m+H_m^\dagger$, and $T_{M}^{c1}$, $B_{M}^{c1}$ correspond to negative-energy solution states (and similarly   for opposite spin components)
\begin{table}[ht]
\noindent \begin{centering}
\begin{tabular}{|c|c|c|c|}
\hline
  massive quarks & $H_m^h$ & $Q$ & $\frac{3i}{2}B\gamma^{1}\gamma^{2}$\tabularnewline
\hline
\hline
$ T_{M}^1 =\frac{1}{\sqrt{2}} ({T_L^1}+{T_R^1} )$ & $m_t$ & $2/3$ & $ 1/2$\tabularnewline
\hline
$B_{M}^1=\frac{1}{\sqrt{2}} ({B_L^1}-{B_R^1} )$ & $m_b$ & $-1/3$ & $ 1/2$\tabularnewline
\hline
$T_{M}^{c1} =\frac{1}{\sqrt{2}} (T_{L}^1-T_{R}^1 )$ & $-m_t$ & $2/3$ & $ 1/2$\tabularnewline
\hline
$B_{M}^{c1}=\frac{1}{\sqrt{2}}({B_L^1}+{B_R^1})$ & $-m_b$ & $-1/3$ & $ 1/2$\tabularnewline
\hline
\end{tabular}
\par\end{centering} \label{Tmassivequark}
\caption{Massive quark  eigenstates of $H_m^h$ given after Eq. \ref{MassiveEigenvalues}}.
\end{table}
and  Eq. \ref{MassiveEigenvalues} justifies  the $m_t$ and  $m_b$ mass interpretation.


%


Under the assumption of a   {\it single} mass-producing field  operator, we match
a reparameterized  $H_n$ in  Eq. \ref{ZmassEqEx} that gives   the  Z  mass,  to   the fermion-mass  term  $H_m$, in Eq.  \ref{MassiveEigenvalues},   resulting in  ${\sqrt{2}H_{n}= }H_{m}$; a  multiplet structure is suggested. In other words,   the operator identification  derives from their mass eigenvalues,  expressed schematically as $|\langle Z |{\sqrt{2}} H_{n}   | Z  \rangle|^2 =m_Z^ 2$ and
  $\langle t |  H_{m} +H_{m}^\dagger | t  \rangle  =m_t $,   and the proportionality constant is  derived  accordingly.
 In this association,   the  simple real-field $Z_\mu(x)$    nature  justifies its use (similarly for each  $W_\mu^i (x) )$,  as opposed to the       complex $W_\mu^\pm(x)$.
 Similarly, Eq. \ref{EquivSV} is chosen over Eq. \ref{EquiSVHer}, as the latter adds the Higgs conjugate representation, unlike the SM.
Thus, the  vacuum expectation value
reproduces  the parameterization in Eq. \ref{HiggsComponentsYm1}, and   identifies  $\chi_t$, $  \chi_b$   as Yukawa parameters: \begin{eqnarray}
\label {YukawaParameters} \ \ \ \  \chi_t= m_t/\frac{v}{\sqrt{2}}  ,  \ \ \  \chi_b= m_b/\frac{v}{\sqrt{2}} .\end{eqnarray} The same argument can be made using the second scalar form in  Eq. \ref{EquiSVHer}, as it also leads to Eq. \ref{ZmassEqEx}. This results in
\begin{eqnarray}
\label {BilinHiggsFerCorrex}
-{\cal L}_{SF}={\rm tr}
 {\sqrt{2}}   [     \Psi_{tR} ^\dagger(x) {\bf H}_{af}(x) { \boldmathPsi}_{qL}(x) +
  {  \boldmathPsi}_{qL}  ^\dagger(x) {\bf H}_{af}(x)  \Psi_{bR} (x)]   +\{ hc \}.
\end{eqnarray}
 Using Eq. \ref{HiggsNormVacuumExpRR},
 we obtain the  relation for the $ t$, $ b$ quark masses
 \begin{eqnarray}
\label  {HiggsNormVacuumEx}
 (|a|^2+|f|^2)v^2/2  = |m_t|^2+|m_b|^2  =v^2/2.
  \end{eqnarray}
  The  commutator arrangement in Eq. \ref{ZmassEq} is used  in the above comparison;  as it  is set on the demand of a  normalized scalar,  the argument strengthens   on the use of the same Z operator  acting on fermions in Eq. \ref{BilinNoTracElectroweak}.
The coefficient matching
 in ${\cal L}_{SF}$    derives from the  underlying freedom of choice in    ${\cal L}_{SV}$,       and, in turn, from the underlying three-parameter $\tau_3$-${\cal C}$ symmetry that can be equally implemented in the spin basis. Looking at the matrix structure, the $\gamma_0$ operator within $H_m$   makes
it  a rank-2 reducible-representation operator, as expressed in  Eq.  \ref{HiggsComponentsYm1} and can be read in Eq. \ref{LSVHiggschi}; indeed,  $H_m$   connects two fermion spin polarizations, but  hits a single W state's components twice as $H_m$ duplicates the scalar representations, requiring the $\frac{1}{\sqrt{2}}$ normalization factor. In yet another interpretation, this relation is obtained from   the  normalization restriction in the Yukawa term  in Eq. \ref {HiggsNormVacuumExpRR},
dividing out the energy scale set by the vacuum expectation. To the extent that these arguments rely
on a common metric vector space,  they are     geometric.

 Equation \ref{HiggsNormVacuumEx} assumes the parity-conserving
condition,    constraining   the   quark masses.\footnote{We neglect  t-b mixing as the CKM matrix is  nearly diagonal\cite{ParticleData},
confirming  this method can be applied here.}   For maximal hierarchy\cite{BesproRicardo}, with  $a$, $f$ dependence
   on  {\it one }  comparable large scale O($a ) \simeq$O($ f$),   ($m_b \ll m_t$,)  we get $\frac{1}{\sqrt{ 2}}v\simeq  173.95   $, for $v=246$  GeV,   $m_b =0$;     alternatively, the quark-b mass input predicts  the top quark-mass as     $ m_t = \sqrt{v^2/2-m_b^2 }\simeq 173.90  $ GeV, for\cite{ParticleData}  $m_b = 4$ GeV (while renormalization effects give\cite{quarkrunmass} $m_b(m_t)\sim 2$ GeV.)
    These two calculations are consistent with   the measured  top pole mass\cite{ParticleData} $\bar m_t=173.21\pm 0.51\pm0.71 $  GeV, where systematic and statistic errors are quoted, respectively.   Future precision improvements will  test the limits of this tree-level calculation, with view of the bottom-quark influence.

\section{Extended quark-mass relation }
 We  place the heavy-quark mass relation in Eq. \ref{HiggsNormVacuumEx} in the larger  SM context, and  argue for a plausible  generalization for all quarks, based on it.  For these purposes, we first derive some SM field properties  using  the spin basis,  assuming  they can be also derived  within the  conventional SM  basis,  given their  equivalent application. Needless to say,
 we
 demand consistency  with   the
  SM,  and with experiment. At the Section's end we  identify some underlaying general assumptions.

   Thus, we concentrate on the SM three-generation subset of the (7+1)-d model\cite{Romero},  as can be effected by the  Yukawa  operators in Eq. \ref{YukawaOp}.
Eq. \ref{HiggsNormVacuumEx} uses that the same single-scalar operator  acts on the  fermions and    the vector bosons:
 such an operator is reproduced in the  $SV$  and $SF$ terms, as the $SV$ term admits a basis expression that applies the associated ${ \cal C }$-symmetries in Section 4.
This  connection implies  the equivalent expression  that  can be read from the Appendices,    \begin{eqnarray}
\label  {YukawaSVextend} {\cal L}_{SV}= {|\chi_t|^2  {\cal L}_{SVu}+|\chi_b|^2 {\cal L}_{SVd}},  \end{eqnarray}
which shows separation    of quark $i=$u- and d-type  ${\cal L}_{SVi}$  components,  depending on scalars,  and no mixing among them.
 We focus on the mass-generating scalar elements corresponding to the neutral $H_t^0$, $H_b^0$, from Eqs. \ref{DefHtHb}, \ref{HiggsComponentsYm1}, and their hermitian conjugates. As mass relations are considered,
we assume fields  after the Higgs mechanism is applied.

In particular, a connection emerges between  the normalized   bilinear Higgs term that gives masses to the vector bosons, as the Z mass in Eq. \ref{ZmassEqEx},  and the fermions.
\begin{eqnarray}
\label  {YukawaSVWithWithoutf}
 \frac{1}{2}{\rm tr}      {   H_m } ^\dagger {   H_m} & = &2 {\rm tr}  [  (  H_m { T}^\alpha_{L}  {T_R^\alpha}^\dagger ) ^\dagger  {   H_m} { T}^\alpha_{L}  {T_R^\alpha}^\dagger+
 (  H_m { B}^\alpha_{R}  {B_L^\alpha}^\dagger ) ^\dagger  {   H_m} { B}^\alpha_{R}  {B_L^\alpha}^\dagger]
\\ \nonumber
&  =& 2 {\rm tr}  [      H_m^\dagger {   H_m} { T}^\alpha_{L}  {T_R^\alpha}^\dagger {{ T}^\alpha_{R}}  {T_L^\alpha}^\dagger  +
     H_m^\dagger {   H_m} { B}^\alpha_{R}  {B_L^\alpha}^\dagger {{ B}^\alpha_{L}}  {B_R^\alpha}^\dagger] = v^2 (\chi_t^2+\chi_b ^2) ,  \end{eqnarray}
where  $H_m$ is defined in Eq. \ref{HmVacuum},
      ${ T}^\alpha_{R}$,  ${ T}^\alpha_{L}$, ${ B}^\alpha_{R}$,  ${ B}^\alpha_{L}$,     are quarks     at rest, defined in     Table \ref{tab:tableqbquarks}, $\chi_t$,  $\chi_b$   are
    $H^0_t$,   $H^0_b$  coefficients, as given in Eq.   \ref{HiggsComponentsYm1},
      in the second equality we use the trace property, and the third expresses the $H_m$ normalization condition.
     Factor 2 comes as only one spin fermion component is used.
       Thus, ${\cal L}_{SV}$ elements can be written as a sum of  inner products between Yukawa and SM scalar components.
This relation derives from the  projective nature of Higgs  normal and dual  terms,
   accompanied by a   fermion chirality operator      in   ${\bf H}_t (x)$, $  {\bf   H}_b(x) $ in Eq. \ref{DefHtHb}.

 Eq. \ref{YukawaSVWithWithoutf} can also be understood from the substitutions in $ {\rm tr}      {   H_m } ^\dagger {   H_m}$
   \begin{eqnarray}
\label  {TransfoHFerBa}
  {   H}^0_{t}& \rightarrow   &   {    H}^0_{t}    { T}^\alpha_{L}  {T_R^\alpha}^\dagger = \frac{1}{\chi_t}y^U_{qt} {    H}^0_{t}    { T}^\alpha_{L}  Y^F_{qt}  {T_R^\alpha}^\dagger     \\ \nonumber
 {   H}^0_{b}& \rightarrow    &   {   H}^0_{b}    B_{R}^\alpha {{   B}^\alpha_{L}}^\dagger = \frac{1}{\chi_b}y^D_{bq}{   H}^0_{b}  B_{R}^\alpha Y^F_{bq}   {{   B}^\alpha_{L}}^\dagger  ,  \end{eqnarray}
with terms extracted from  $ {\cal L}_{SF}$ in Eq. \ref {BilinHiggsFer}, using the trace permutation property.
The identity in each substitution provides the
   link to the t, b Yukawa constants
for the    $q=tb$  doublet,  $t,b$ singlet  cases.  The arguments leading to the mass relation in Eq.  \ref{HiggsNormVacuumEx} imply  $y^U_{qt}=\chi_t$,  $y^D_{bq}=\chi_b$, as given in Eq.  \ref{YukawaParameters}, namely, a  diagonal mass  basis is assumed.

Since one can pick any   fermion generation on Tables  \ref{tab:masslesslefthanded}, \ref{tab:masslessrighthanded},
 the interpretation of the  $\chi_t$, $\chi_b$ coefficients as Yukawa constants within the $SV$ term leads to a generalization to other families and non-diagonal Yukawa elements.
We  now consider  the extension of ${\cal L } _{SF}$    in  Eqs. \ref{BilinHiggsFer} and  \ref{BilinHiggsFerCorrex}  with
  a  fermion expansion that uses   all Yukawa coefficients,
\begin{eqnarray}
\label {BilinHiggsGenYuka}
-{\cal L}_{SFT}=  {\rm tr}
   [ \sum_{iq} y^U_{iq}  \Psi_{iR}^\dagger(x) {\bf   H}_{t}(x)   { \boldmathPsi}_{qL}(x) Y^F_{iq} +\sum_{jq}
y^D_{qj} {  \boldmathPsi}_{qL}  ^\dagger(x) {\bf  H}_{b}(x)   \Psi_{jR} (x) Y^F_{qj}]    +\{ hc \},
\end{eqnarray}    where   the Yukawa operators $Y^F$  from Eq. \ref{YukawaOp} are necessary to connect  the u- and d-type quark fields  defined in Eq. \ref{quarkfields}, and  $y^U_{qi}$,   $y^D_{qj}$ are  Yukawa coefficients, with   the   up, down, charm  and strange quarks, also included,
relabelling singlets ${i}=u, c, t$,  ${j}= d, s, b,$  and doublets  ${q}= ud, cs, tb$.



The  allowed Yukawa terms, diagonal and mixed, can be included using all combinations of  a 3-generation  set of  normalized fermions on Table \ref{tab:masslesslefthanded}, where a projection operator as in Eq. \ref{ProjeFla} is  applied.
       We evaluate the trace of bilinear
  $F_U=\frac{1}{\chi_t} y^U_{qi}  U^\alpha_{qL}  Y^F_{qi}  {  U_{iR}^\alpha}^\dagger$,
 $F_D=\frac{1}{\chi_b} y^D_{jq'}  D^\alpha_{jR}  Y^F_{jq'}   { D_{q'L}^\alpha}^\dagger$
  terms with ${\cal L}_{SV}$ components,   extending Eq. \ref{YukawaSVWithWithoutf},  producing

 \begin{eqnarray}
\label  {TransfoHHp}  & & 2 {\rm tr}  [ (  H_m F_U    )  ^\dagger     H_m  F_U+    (H_m F_D)^\dagger (  H_m F_D    )  ]
\\ \nonumber
 &=  &  \frac{1}{2}  v^2(  |y^U_{qi}|^2   {\rm tr}  {  H^0  }^\dagger_{t}  {   H }^0_{t} +
  |y^D_{jq'}|^2    {\rm tr} {   H^0  } ^\dagger_{b}   {    H  }^0_{b}) \\ \nonumber
   &= &v^2(|y^U_{qi}|^2    +
   |y^D_{jq'}|^2 ),    \end{eqnarray}
   which may be also obtained by the substitution of the associated scalar coefficients in   bilinear neutral Higgs terms
      $ {\rm tr}      {   H_m } ^\dagger {   H_m}$
 \begin{eqnarray}\label  {TransfoHFer}
  {    H}_{t}(x)&\rightarrow &{   H}^0_{t}    F_U ={   H}^0_{t}  \frac{1}{\chi_t} y^U_{qi}  U^\alpha_{qL}  Y^F_{qi}  {  U_{iR}^\alpha}^\dagger \\ \nonumber
 {   H}_{b}(x)&\rightarrow  & {   H}^0_{b}   F_D = {   H}^0_{b}   \frac{1}{\chi_b} y^D_{jq'}  D^\alpha_{jR}  Y^F_{jq'}   { D_{q'L}^\alpha}^\dagger. \end{eqnarray}
   The correspondence of ${\cal L } _{SF}$  in  Eqs.  \ref{BilinHiggsFer} and  \ref{BilinHiggsFerCorrex} to ${\cal L}_{SFT}$ in    Eq. \ref{BilinHiggsGenYuka}   induces the sum of  square mass-matrix elements in Eq. \ref {TransfoHHp}, which is equal  (given the property ${\rm tr} M^{\dagger} M ={\rm tr} M^{\prime\dagger} M'$, $M$ a matrix, $M'$ its diagonal form) to the  sum over the        square masses, \begin{eqnarray}
\label  {MassSumRuleGen}       v^2 (\sum_{qi}   |y^U_{qi}|^2 +    \sum_{qj}
    |y^D_{qj}|^2 )  = 2(\sum_{i}m _i^2+\sum_{j}m _j^2)   .  \end{eqnarray}
 A   generalization  with  such a sum is induced,  similar to relation Eq.    \ref{HiggsNormVacuumEx},  with the  Higgs normalization condition, Eq. \ref{HiggsNormVacuumExpRR}. Since
    Eq. \ref {TransfoHHp} maintains the same structure as Eq.  \ref{YukawaSVWithWithoutf}, following the
  generalization of ${\cal L}_{SF}$    to ${\cal L}_{SFT}$,
\begin{eqnarray}
\label  {MassSumRule}
m_t^2+m_c^2+m_u^2+m_b^2+m_s^2+m_d^2=v^2/2.
  \end{eqnarray}
Implicitly,   we used the $SV$-fermion symmetry, namely, no  fermion  preference.
  With today's uncertainties in the quark-mass values, this relation is phenomenologically consistent with
  Eq. \ref{HiggsNormVacuumEx}, as the  same maximal hierarchy or quark b-mass input argument follows,  and  the rest of the quarks have comparably negligible masses. As this  relation is independent of the mass diagonalization matrix, it is also of the CKM matrix\cite{KobayashiMaskawa}.




The two quark-mass conditions in  Eqs. \ref{HiggsNormVacuumEx} and  \ref{MassSumRule} are  interpreted.  This paper shows SM features support a boson and fermion connection leading to the t,b quark mass condition in  Eq. \ref{HiggsNormVacuumEx}.   If only  such quarks belong in the same
class as the other  massive SM bosons,  a    different mass-generating mechanism  is expected for the other fermions; one   concludes   that they    are not affected by such dynamics, as  their masses are comparably negligible.
On the other hand, if there is a common dynamics, as suggested by the similar fermion-boson inner product,  the  all-quark condition Eq. \ref{MassSumRule} applies, given the fermion symmetry, and
   the structure similarity between Eq.   \ref{YukawaSVWithWithoutf}   and   Eq.  \ref{TransfoHHp}.

 Initial fermion states within the 3-generation  set  for ${\cal L}_{SFT} $ in Eq.    \ref{BilinHiggsGenYuka}    remain  within such a subspace, given
the commuting property of the projection operator $Y^{F4}$ in Eq. \ref{ProjeFla} with  baryon-number, Lorentz,  gauge   and mass operators ($B$, $B \sigma_{\mu\nu}$  $I_i$, $Y_{o}$, $\boldmathphi_{i}$, $i= 1,2$.)   In other words, within the 3-generation subset of states,    the substitution  $Y^{ F}_{ij}\rightarrow {Y^{ F4}}^\dagger Y^{ F}_{ij} {Y^{F4}}$ in ${\cal L}_{SFT} $  is valid.
This implies that no operator will  connect the initial fermions outside the 3 generations. So is the case for the 3-generation extension of ${\cal L}_{FV} $ in Eq. \ref {BilinNoTracElectroweak},  requiring a  sum over the (electroweak) flavors.
We conclude the 3-generation
  spin-basis projection   consistently describes the SM.


 By
construction,  Eqs. \ref{TransfoHFerBa},  \ref{TransfoHFer} imply masses represent O($m_q/m_t$) corrections. This is also the order of the  Hamiltonian needed to obtain the other fermion masses.
 More assumptions are necessary to get further information on masses, and CKM matrix elements. For example,  hierarchy  arguments on the masses'  order of magnitude difference   were derived\cite{Romero}    that  explain how the associated W,Z,t,b, large scale  mostly cancels for the other fermions  at the vertical level (within a  doublet) and horizontal level (between families).
This leads to a consistent description    in which such    mechanisms  coexist  with the Higgs-generated one.
While we produce above further consistency arguments for their parameters, more stringent constraints from the (7+1)-d will be tested elsewhere. Other arguments leading to hierarchy exist as textures\cite{fermionMassReview}.


  We conclude   Yukawa  coefficients, contained in rest  fermions     as    a device,   connect  to    bilinear scalar combinations containing mass-generating Higgs terms in ${\cal L}_{SV }$,  keeping  the Lorentz or gauge structure of $SV$ unmodified, and ultimately consistently with the SM. We show above ${\cal L}_{SFT }$ in Eq.
  \ref{BilinHiggsGenYuka} induces a generalized    sum rule for the square quark masses in Eq. \ref{MassSumRule}.  The latter is a plausible extension of Eq.  \ref{HiggsNormVacuumEx}, based on a subset of  ${\cal L}_{SV }$ terms, after the Higgs mechanism.
   The same type of argument can be made for
  leptons, but given their smaller   masses, their influence will be lesser, while similar conditions as in Eq. \ref{YukawaSVWithWithoutf} will also lead to
  PNRS     matrix\cite{PNRS}  independence.

\section{Conclusions and Outlook}


In summary, the formalism used places fields on a   basis  that
simultaneously  contains  SM  bosons  and  fermions. $SV$ and $SF$ terms are linked through the mass rendering of the  scalar operator within them,
   using    the  electroweak $SV$ vertex   independence
of its components acting on different fermion-doublet elements,
  implicitly expected, but which we  now expose. Supporting  a
  SM prediction of a unique scalar,   input     from  the  normalized scalar-vector
vertex,  and  the  mass-parameter interpretation in  the   $SF$ vertex, relates $v$ and $m_t$, cf Eq. \ref{HiggsNormVacuumEx},  the main result in the paper.  The
  same relation can be argued  by considering the scalar operator's  matrix rank, or assuming normalized Yukawa components.
   Based on  chiral properties,  the same Higgs-operator rule,   and  a correspondence between    fermion-boson inner products and Yukawa terms,
       a plausible   extended sum rule   for the fermion square masses is proposed, given in  Eq. \ref{MassSumRule}. Both relations are consistent   with  the SM, given today's particle-mass uncertainties.
  We conclude the spin basis is a useful platform to obtain, within the SM, the quark-mass electroweak relations.

The  central argument input can be also read   when  $V$ terms
in   ${\cal L}_{FV}$, attached with the projector $L_5$ in Eq. \ref{projeRL},  are carried   into the intermediate    ${\cal L}_{SV}$  chiral   version in Eq. \ref{LSVHiggschi}     and,
 after the 1/2 factor cancellation in its mass component, relate  to  $F$ terms in the Yukawa ${\cal L}_{SF}$.
The    spin-basis   gives  it further support as it classifies discrete degrees and produces SM  features.
Thus,
 the matrix space restricts representations, in turn, exhausting the  space;
   electroweak $V$ fields  belong to the  adjoint, and $S$, $F$ fields to  the fundamental representations.
Additionally,
  the    chiral property    in  the   $FV$  electroweak term, associated to $V$,   translates naturally to the $SV$   interaction  components. Normalized  fields define
  the   Lagrangian terms, setting  the trace coefficient, and
 the stage for the ${\cal L}_{SV}$, ${\cal L}_{SF}$ comparison.
    In the spin-basis context, the    $S$ field's    chiral property is  nominal, but consistent, as
 ${\cal L}_{SV}$       contains  the $L_5$ projector from $V$,
and  within ${\cal L}_{SF}$,     $S$  acts on chiral fermion components.



The scalar operator  acting   on vectors and fermions links their matrix elements, connecting parameters.
  The    particles'   simultaneous  participation in mass generation through   the   Higgs mechanism and   related  SM  vertices,
  with assigned representations, implies   a  description  with common dynamics, and at a  given energy scale,
     already at the classical level, and
         suggests   fields belong in a  multiplet,     supporting a  common-origin unification assumption\cite{BesproRicardo}.

It follows that  the arguments provide  a geometric approach  to address   problems  as the electroweak-symmetry breaking origin.
The formalism facilitates the   fields'  composite description,  as boson degrees of freedom may be written in terms of two fermions'.
Expansions in such fields   may be useful, independently of whether compositeness is  physical or only a  device.


Naturalness is hinted at in  the  $ {\boldmathphi }_1$,  ${\boldmathphi  }_2$    associated single scale, which produces a hierarchy effect\cite{BesproRicardo}.
 Thus, while this  symmetry-breaking   effect applies  for heavy-quark masses,
   it could be valid also horizontally between generations in accordance  with the   fermions' low masses.
While here we considered the top-quark mass,  the other fermions, besides the b-quark, may be included in this scheme, namely, considering bilinear  fermion components for scalar particles,
but they will have  little influence on this result, as their $SF$  interaction is   proportional to their masses.

 As the spin basis connects the vector and quark sectors, constraints may be derived for      SM extensions
   as supersymmetry\cite{supersymmetry},    composite models that require    dynamical  symmetry breaking\cite{NambuJona} as
 technicolor\cite{technicolor}
   or, in an extension of  such models,
 top and  bottom   quarks\cite{Bardeen} that  conform   condensate-producing massive particles.



Besides  the fields' spin representation  connecting the scalar operator in two vertices, it
highlights chiral components of     particles and   interactions
 that   maintain  their SM equivalence.
  Indeed,  we showed two   such valid   chiral and non-chiral scalar bases for the   $SV$ Lagrangian.
  This  freedom  could be clarified in other vertices, as with a SM extension with additional scalar degrees of freedom, whereas in this paper, we considered only their SM projection.

{\bf Acknowledgements}
 The
authors acknowledge  support from  DGAPA-UNAM through project IN112916, and discussions with A. Ayala.

\appendix{}
\section*{Appendix 1: Fermion-vector   ${\cal L}_{FV}$ fermion-scalar $-{\cal L}_{SF}$ Lagrangians  }


In this  Appendix we    show  the  Lagrangians'   equivalence    in the conventional and spin  bases by considering  explicit   expressions with accompanying wave functions (or fields).
With hindsight, we use the same   Lagrangian   label in both bases.

First, we use an iterative procedure[16] to obtain a (7+1)-d  $\gamma^\mu$ representation.
Starting with the Pauli matrices $\sigma^{1}$, $\sigma^{2}$ and
$\sigma^{3}$, we get the $(3+1)$-d representation

\begin{eqnarray}
\begin{array}{cl}
\alpha^{0}=\sigma^{1}\otimes\sigma^{3} & \alpha^{1}=-i\sigma^{2}\otimes\sigma^{3},\\
\alpha^{2}=I_{2}\otimes i\sigma^{1} & \alpha^{3}=I_{2}\otimes i\sigma^{2};
\end{array}\label{eq:A1}
\end{eqnarray}

\noindent then, the $(5+1)$-d  representation

\begin{equation}
\begin{array}{ll}
\beta^{0}=\alpha^{0}\otimes\sigma^{3} & \beta^{1}=\alpha^{1}\otimes\sigma^{3},\\
\beta^{2}=\alpha^{2}\otimes\sigma^{3} & \beta^{3}=\alpha^{3}\otimes\sigma^{3},\\
\beta^{5}=I_{4}\otimes i\sigma^{1} & \beta^{6}=I_{4}\otimes i\sigma^{2},
\end{array}\label{eq:A2}
\end{equation}

\noindent and finally, the $(7+1)$-d representation

\begin{equation}
\begin{array}{ll}
\gamma^{0}=\beta^{0}\otimes\sigma^{3} & \gamma^{1}=\beta^{1}\otimes\sigma^{3},\\
\gamma^{2}=\beta^{2}\otimes\sigma^{3} & \gamma^{3}=\beta^{3}\otimes\sigma^{3},\\
\gamma^{5}=\beta^{5}\otimes\sigma^{3} & \gamma^{6}=\beta^{6}\otimes\sigma^{3},\\
\gamma^{7}=I_{8}\otimes i\sigma^{1} & \gamma^{8}=I_{8}\otimes i\sigma^{2}.
\end{array}\label{eq:A3}
\end{equation}


  The commuting property of the Lorentz and scalar symmetry operators implies that they can be represented as a tensor product.   To compare with the spin-space basis,
    we write    the conventional-basis generators  as  tensor products,   choosing the (7+1)-d space to represent them;  thus
  the  spin-1/2   and SU(2)$_L$ terms,  expressed by the $4\times4$  Clifford  basis, and Pauli matrices, respectively,  generalize to, e.g.,   $ \tau_3\bigotimes 1_{s 2\times2}\sim I^3 $
  and  $1_{w 2\times2}\bigotimes [  \frac{i}{2}P_L \gamma_1\gamma_2]_{2\times 2}   \sim  \frac{i}{2}P_L \gamma_1\gamma_2 $, with $\tau_3$  the 3-Pauli matrix, and corresponding
  spin and weak isospin unit operators  $1_{s 2\times2}$, $1_{w 2\times2}$, respectively.

Similarly, states in the conventional basis can be obtained that are    represented in (7+1)-d space.
For example,    a left-handed (L),  spin-1/2 polarization (1),   top (T),     state
$ |  L 1  T   \rangle $ satisfies
$\frac{1}{2}(1-\tilde \gamma_5)| L 1 T\rangle = -| L 1 T\rangle  , $  \  \
$ i\frac{1}{2}P_L \gamma_1\gamma_2| L 1 T\rangle  =\frac{1}{2}| L 1 T\rangle, $ \ \
$ I^3| L 1 T\rangle =\frac{1}{2} | L 1 T\rangle $.

Eq. 36 in the paper implies spinors are labeled by the $4\times 4$  spin  operator in the Dirac representation $ \frac{i}{2} \gamma_1\gamma_2 $, and
    the weak SU(L)$_L$ $\tau_3$ component. While most of the results in the paper are representation-independent, a unitary transformation may be applied to the (7+1)-d matrices to show the conventional-basis
description used in Eq. 36 in the paper. Indeed, $    \frac{i}{2}P_L \gamma_1\gamma_2$ has a Dirac   form with the unitary transformation
$\gamma ^\mu_D=U_D^\dagger  \gamma^ \mu   U_D$ with $U_D= \frac{1}{\sqrt{2}}(1-\gamma^1\gamma^3) \gamma^0  \gamma^2 \gamma^ 3 $ (actually, it exchanges $\gamma^1$ and $\gamma^3$.)
and  $ |  L 1 T\rangle $ is represented, after the unitary transformation $U_D^\dagger$,  by

\noindent $ (0, 0, 0, 0, 0, 0, 0, 0, 0, 0, 0, 0, 0, 0, i, 0).$
Next, we write  all the  conventional-basis states  in this basis, and their association to the spin-extended basis states, with corresponding    quantum numbers (notation used  in Table 4 and Ref. 7, written inbetween):

 \[ (0, 0, 0, 0, 0, 0, 0, 0, 0, 0, 0, 0, 0, 0, i, 0) \leftrightarrow T  ^1{}_{ { L}},U  ^1{}_{ { L} 1}  \]
  \[ \nonumber (0, 0, -i, 0, 0, 0, 0, 0, 0, 0, 0, 0, 0, 0, 0, 0) \leftrightarrow T  ^2{}_{ { L}} , U  ^2{}_{ { L} 1}   \]
 \[  (0, 0, 0, 0, 0, 0, -i, 0, 0, 0, 0, 0, 0, 0, 0, 0) \leftrightarrow T  ^1{}_{ { R}} , U  ^1{}_{ { R} 1}  \]
  \[ \nonumber (0, 0, 0, 0, 0, 0, 0, 0, 0, 0, i, 0, 0, 0, 0, 0) \leftrightarrow T  ^2{}_{ { R}} ,U  ^2{}_{ { R} 1}  \]
  \[ \nonumber  ({0, 0, 0, 0, 0, 0, 0, 0, 0, 0, 0, 0, 0, 0, 0, -i}) \leftrightarrow B  ^1{}_{ { L}},D  ^1{}_{ { L} 1}  \]
 \[  (0, 0, 0, i, 0, 0, 0, 0, 0, 0, 0, 0, 0, 0, 0, 0) \leftrightarrow B  ^2{}_{ { L}}, D  ^2{}_{ { L} 1}  \]
  \[ \nonumber ({ 0, 0, 0, 0, 0, 0, 0, i, 0, 0, 0, 0, 0, 0, 0, 0}) \leftrightarrow B  ^1{}_{ { R}} , D  ^1{}_{ { R} 1}  \]
   \[  (0, 0, 0, 0, 0, 0, 0, 0, 0, 0, 0, -i, 0, 0, 0, 0) \leftrightarrow B  ^2{}_{ { R}} , D  ^2{}_{ { R} 1},   \]
where the spin-basis states are shown {\it in
extenso} in Tables 2 and 3.

 For the fermion wave functions $ \psi_{qh}^\alpha(x)  $, we use  polar coordinates, where the conventional and spin terms contain,
 respectively, $\uppsi^\alpha{}   _{qh}(x) \exp{[i p^\alpha{}_{qh}(x)]} \leftrightarrow
 \uppsi^\alpha{}_{qh}(x) \exp{[i c^\alpha{}_{qh}(x)]},$   for quarks  $q=t,b$,  with spin components  $\alpha=1,2$, and chirality   $h=L,R$.
  The magnitude part can be shown to be the same for both cases, as can be derived by comparing, e. g., the mass term. The vectors $W_\mu^a(x)$,
 $B_\mu(x)$  are real fields.

The phases appear in each term in both bases. For example, for the conventional basis and for the two  polarizations   $     t_{L}(x)=\left(\begin{array}{lcr}
     \psi_{tL}^1(x) \\
  \psi_{tL}^2(x)
 \end{array}\right)$  within the   left-handed        hypercharge $Y =1/3$  SU(2)$_L$-doublet,
  we use the association
$t_{L}(x)\rightarrow \uppsi^1{} _{tL}(x)  \exp{[i p^1{}_{tL}(x)]} U_D \left(\begin{array}{lcr}
  0\\  0\\ 0\\ 0\\ 0\\ 0\\ 0\\ 0\\ 0\\ 0\\ 0\\ 0\\ 0\\ 0\\ i\\ 0
 \end{array}\right)+\uppsi^2{} _{tL}(x)  \exp{[i p^2{}_{tL}(x)]}U_D\left(\begin{array}{lcr}
  0\\  0\\ -i\\ 0\\ 0\\ 0\\ 0\\ 0\\ 0\\ 0\\ 0\\ 0\\ 0\\ 0\\ 0\\ 0
\end{array}\right),$ with  $U_D$   applied to transform back from the Dirac representation, and we used the terms in Eqs. (4) and (5) in this Appendix;
for  the spin basis,
${ \Psi_{tL} }(x)=  \uppsi^1{} _{tL}(x)  \exp{[i c^1{}_{tL}(x)]} T^1_L+ \uppsi^2{} _{tL}  (x)  \exp{[i c^2{}_{tL}(x)]} T^2_L $.

 The  Lagrangians' identity is shown, by checking that the same terms are reproduced in both bases, and finding independent  constant phases that connect the two representations.
In the following, we present   the  fermion-vector  ${\cal L}_{FV}$ Lagrangian components:    interactive (weak and hypercharge),   kinetic; also the fermion-scalar  (Yukawa)   ${\cal L}_{SF}$    Lagrangian. The subtitle contains the two-basis Lagrangian expressions in  a concise notation, and then   one component is given in an expanded form;     the equations that link the phases in  the two representations are written as they derive from the terms.

\subsection {Weak. $\bar {  \bf q }_L(x)   \frac{1}{2}g\tau^a  W_\mu^a(x)  \gamma^\mu  { \bf q}_L (x)
\leftrightarrow$ $ {\rm tr } \{  {  \boldmathPsi_{qL}  }(x)  ^\dagger    g I_a  W_\mu^a(x)  \gamma^0 \gamma^\mu \boldmathPsi_{qL} (x) \}  $ }

\[
\frac{g}{2} \left(\left(W^3{}_3(x)-W^3{}_0(x)\right) \uppsi  ^1{}_{{bL}}(x){}^2-2 \left(\left(\cos
   \left(p^1{}_{{bL}}(x)-p^1{}_{{tL}}(x)\right) W^1{}_0(x)\\
   -\cos \left(p^1{}_{{bL}}(x)-p^1{}_{{tL}}(x)\right) W^1{}_3(x)+\sin
   \left(p^1{}_{{bL}}(x)-p^1{}_{{tL}}(x)\right) \left(W^2{}_0(x)-W^2{}_3(x)\right)\right) \uppsi  ^1{}_{{tL}}(x)+\left(\sin
   \left(p^1{}_{{bL}}(x)-p^2{}_{{bL}}(x)\right) W^3{}_2(x)\\
   -\cos \left(p^1{}_{{bL}}(x)-p^2{}_{{bL}}(x)\right) W^3{}_1(x)\right) \uppsi
   ^2{}_{{bL}}(x)
   -\left(\cos \left(p^1{}_{{bL}}(x)-p^2{}_{{tL}}(x)\right) W^1{}_1(x)
   -\sin
   \left(p^1{}_{{bL}}(x)-p^2{}_{{tL}}(x)\right) W^1{}_2(x)\\
   +\sin \left(p^1{}_{{bL}}(x)-p^2{}_{{tL}}(x)\right) W^2{}_1(x)+\cos
   \left(p^1{}_{{bL}}(x)-p^2{}_{{tL}}(x)\right) W^2{}_2(x)\right) \uppsi  ^2{}_{{tL}}(x)\right) \uppsi  ^1{}_{{bL}}(x)+W^3{}_0(x) \uppsi
   ^1{}_{{tL}}(x){}^2-W^3{}_3(x) \uppsi  ^1{}_{{tL}}(x){}^2-\left(W^3{}_0(x)+W^3{}_3(x)\right) \uppsi  ^2{}_{{bL}}(x){}^2+W^3{}_0(x) \uppsi
   ^2{}_{{tL}}(x){}^2+W^3{}_3(x) \uppsi  ^2{}_{{tL}}(x){}^2-2 \cos \left(p^1{}_{{tL}}(x)-p^2{}_{{tL}}(x)\right) W^3{}_1(x) \uppsi
   ^1{}_{{tL}}(x) \uppsi  ^2{}_{{tL}}(x)+2 \sin \left(p^1{}_{{tL}}(x)-p^2{}_{{tL}}(x)\right) W^3{}_2(x) \uppsi  ^1{}_{{tL}}(x) \uppsi
   ^2{}_{{tL}}(x)-2 \uppsi  ^2{}_{{bL}}(x) \left(\left(\cos \left(p^2{}_{{bL}}(x)-p^2{}_{{tL}}(x)\right) W^1{}_0(x)+\cos
   \left(p^2{}_{{bL}}(x)-p^2{}_{{tL}}(x)\right) W^1{}_3(x)\\
   +\sin \left(p^2{}_{{bL}}(x)-p^2{}_{{tL}}(x)\right)
   \left(W^2{}_0(x)+W^2{}_3(x)\right)\right) \uppsi  ^2{}_{{tL}}(x)-\left(\cos \left(p^1{}_{{tL}}(x)-p^2{}_{{bL}}(x)\right) W^1{}_1(x)-\sin
   \left(p^1{}_{{tL}}(x)-p^2{}_{{bL}}(x)\right) W^1{}_2(x)\\
   -\sin \left(p^1{}_{{tL}}(x)-p^2{}_{{bL}}(x)\right) W^2{}_1(x)-\cos
   \left(p^1{}_{{tL}}(x)-p^2{}_{{bL}}(x)\right) W^2{}_2(x)\right) \uppsi  ^1{}_{{tL}}(x)\right)\right).
    \]
 Comparing the corresponding expression in the spin basis, we derive the following phase relations (which retroactively provide such an expression).

\[c^1{}_{ {bL}}(x)= p_{ {W2}}+p^1{}_{  {bL}}(x)+\frac{\pi }{2} \]

\[c^1{}_{  {tL}}(x)= p_{  {W2}}+p^1{}_{  {tL}}(x)+\pi  \]

\[c^2{}_{ {bL}}(x)= p_{  {W3}}+p^2{}_{ {bL}}(x)+\frac{\pi }{2}  \]

\[c^2{}_{  {tL}}(x)= p_{  {W4}}+p^2{}_{  {tL}}(x)+\pi  \]

\[c^2{}_{ {bR}}(x)= p_{  {Z1}}+p^2{}_{  {bR}}(x)  \]

\[c^2{}_{ {tR}}(x)= p_{ {Z2}}+p^2{}_{  {tR}}(x)  \]

\[c^1{}_{ {bR}}(x)= p_{  {Z1}}+p^1{}_{  {bR}}(x)+\pi  \]

\[c^1{}_{ {tR}}(x)= p_{  {Z2}}+p^1{}_{  {tR}}(x)+\pi ,  \]
for arbitrary real constants $ p_{  {W1}}$, $ p_{  {W2}}$, $ p_{  {W3}}$, $ p_{  {W4}}$, $ p_{  {Z1}}$,  $ p_{  {Z2}}$,
requiring the identities
$ p_{  {W1}}= p_{  {W3}}=p_{  {W4}}=p_{  {W2}}$.

   \subsection {Hypercharge. $\bar {  \bf q }_L(x)  \frac{1}{6}g'  B_\mu (x)  \gamma^\mu  { \bf q}_L (x)  +{ \bar   t_R      }(x)      [\frac{2}{3}g'  B_\mu (x) ]    \gamma^\mu    t_{R}(x)
 +{ \bar   b_R  (x)    }      [    -\frac{1}{3}g'  B_\mu (x) ]    \gamma^\mu    b_{R}(x)
\leftrightarrow {\rm tr } \{  {  \boldmathPsi_{qL}  }(x)  ^\dagger  \frac{1}{2}  g' Y_o B_\mu (x)   \gamma^0 \gamma^\mu \boldmathPsi_{qL} (x) +\Psi _{tR } ^\dagger (x)          \frac{1}{2}  g' Y_o B_\mu (x)      \gamma^0 \gamma^\mu     {     \Psi _{tR }}(x)
 +{     \Psi}_{bR}  ^\dagger(x)               \frac{1}{2} g'  Y_o  B_\mu (x)     \gamma^0 \gamma^\mu    { \Psi_{bR} }(x)  \}  $ }
 \[
 \frac{g'}{6} \left(\left(B_0(x)-B_3(x)\right) \uppsi ^1{}_{{bL}}(x){}^2-2 \left(\cos
   \left(p^1{}_{{bL}}(x)-p^2{}_{{bL}}(x)\right) B_1(x)-\sin \left(p^1{}_{{bL}}(x)-p^2{}_{{bL}}(x)\right) B_2(x)\right)
   \uppsi ^2{}_{{bL}}(x) \uppsi ^1{}_{{bL}}(x)-2 B_0(x) \uppsi ^1{}_{{bR}}(x){}^2-2 B_3(x) \uppsi ^1{}_{{bR}}(x){}^2+B_0(x) \uppsi
   ^1{}_{{tL}}(x){}^2-B_3(x) \uppsi ^1{}_{{tL}}(x){}^2+4 B_0(x) \uppsi ^1{}_{{tR}}(x){}^2+4 B_3(x) \uppsi
   ^1{}_{{tR}}(x){}^2+\left(B_0(x)+B_3(x)\right) \uppsi ^2{}_{{bL}}(x){}^2-2 B_0(x) \uppsi ^2{}_{{bR}}(x){}^2+2 B_3(x) \uppsi
   ^2{}_{{bR}}(x){}^2+B_0(x) \uppsi ^2{}_{{tL}}(x){}^2+B_3(x) \uppsi ^2{}_{{tL}}(x){}^2+4 B_0(x) \uppsi ^2{}_{{tR}}(x){}^2-4
   B_3(x) \uppsi ^2{}_{{tR}}(x){}^2+4 \cos \left(p^1{}_{{bR}}(x)-p^2{}_{{bR}}(x)\right) B_1(x) \uppsi ^1{}_{{bR}}(x) \uppsi
   ^2{}_{{bR}}(x)-4 \sin \left(p^1{}_{{bR}}(x)-p^2{}_{{bR}}(x)\right) B_2(x) \uppsi ^1{}_{{bR}}(x) \uppsi
   ^2{}_{{bR}}(x)-2 \cos \left(p^1{}_{{tL}}(x)-p^2{}_{{tL}}(x)\right) B_1(x) \uppsi ^1{}_{{tL}}(x) \uppsi
   ^2{}_{{tL}}(x)+2 \sin \left(p^1{}_{{tL}}(x)-p^2{}_{{tL}}(x)\right) B_2(x) \uppsi ^1{}_{{tL}}(x) \uppsi
   ^2{}_{{tL}}(x)-8 \cos \left(p^1{}_{{tR}}(x)-p^2{}_{{tR}}(x)\right) B_1(x) \uppsi ^1{}_{{tR}}(x) \uppsi
   ^2{}_{{tR}}(x)+8 \sin \left(p^1{}_{{tR}}(x)-p^2{}_{{tR}}(x)\right) B_2(x) \uppsi ^1{}_{{tR}}(x) \uppsi
   ^2{}_{{tR}}(x)\right) .
\]

 \subsection {Kinetic. ${\bar {\bf q }_L}(x)  i\frac{1}{2}\stackrel{\leftrightarrow}{\partial_\mu}    \gamma^\mu  { \bf q}_L (x)     +  { \bar   t_R    }(x)        i\frac{1}{2}\stackrel{\leftrightarrow}{\partial_\mu}      \gamma^\mu    t_{R}(x)
 +{ \bar   b_R  (x)    }       i\frac{1}{2}\stackrel{\leftrightarrow}{\partial_\mu }        \gamma^\mu    b_{R}(x)
\leftrightarrow$ $ {\rm tr }\frac{1}{2} \{ { \boldmathPsi_{qL}   ^\dagger     }(x)         i\stackrel{\leftrightarrow}{\partial_\mu }    \gamma^0  \gamma^\mu   \boldmathPsi_{qL} (x) +
         \Psi _{tR } ^\dagger (x)         i\stackrel{\leftrightarrow}{\partial_\mu}       \gamma^0 \gamma^\mu     {     \Psi _{tR }}(x)
 +{     \Psi}_{bR}  ^\dagger(x)               i\stackrel{\leftrightarrow}{\partial_\mu  }     \gamma^0 \gamma^\mu    { \Psi_{bR} }(x)  \} $ }
 Using the fields' integrability property (belonging to Hilbert space),   integration by parts has been applied to make the  derivative  substitution  $ i{\partial_\mu}\rightarrow i\frac{1}{2}\stackrel{\leftrightarrow}{\partial_\mu}$.

\[
-\uppsi  ^2{}_{{tL}}(x){}^2 \left({   }\partial _zp^2{}_{{tL}}(x) \right)+\uppsi  ^2{}_{{tR}}(x){}^2 \left({
   }\partial _zp^2{}_{{tR}}(x) \right)-2 {\sin }\left[p^1{}_{{bR}}(x)-p^2{}_{{bR}}(x)\right] \uppsi
   ^1{}_{{bR}}(x) \uppsi  ^2{}_{{bR}}(x) \left({   }\partial _yp^1{}_{{bR}}(x) \right)-2
   {\sin }\left[p^1{}_{{bR}}(x)-p^2{}_{{bR}}(x)\right] \uppsi  ^1{}_{{bR}}(x) \uppsi  ^2{}_{{bR}}(x)
   \left({   }\partial _yp^2{}_{{bR}}(x) \right)+2 {\cos }\left[p^1{}_{{bR}}(x)-p^2{}_{{bR}}(x)\right] \uppsi
   ^2{}_{{bR}}(x) \left({   }\partial _y\uppsi  ^1{}_{{bR}}(x)
   \right)+{\cos }\left[p^1{}_{{tL}}(x)-p^2{}_{{tL}}(x)\right] \uppsi  ^2{}_{{tL}}(x) \left({   }\partial
   _y\uppsi  ^1{}_{{tL}}(x) \right)+{\cos }\left[p^1{}_{{tR}}(x)-p^2{}_{{tR}}(x)\right] \uppsi
   ^2{}_{{tR}}(x) \left({   }\partial _y\uppsi  ^1{}_{{tR}}(x) \right)-2
   {\cos }\left[p^1{}_{{bR}}(x)-p^2{}_{{bR}}(x)\right] \uppsi  ^1{}_{{bR}}(x) \left({   }\partial _y\uppsi
   ^2{}_{{bR}}(x) \right)+2 {\cos }\left[p^1{}_{{bR}}(x)-p^2{}_{{bR}}(x)\right] \uppsi  ^1{}_{{bR}}(x)
   \uppsi  ^2{}_{{bR}}(x) \left({   }\partial _xp^1{}_{{bR}}(x) \right)+2
   {\cos }\left[p^1{}_{{bR}}(x)-p^2{}_{{bR}}(x)\right] \uppsi  ^1{}_{{bR}}(x) \uppsi  ^2{}_{{bR}}(x)
   \left({   }\partial _xp^2{}_{{bR}}(x) \right)+2 {\sin }\left[p^1{}_{{bR}}(x)-p^2{}_{{bR}}(x)\right] \uppsi
   ^2{}_{{bR}}(x) \left({   }\partial _x\uppsi  ^1{}_{{bR}}(x)
   \right)+{\sin }\left[p^1{}_{{tL}}(x)-p^2{}_{{tL}}(x)\right] \uppsi  ^2{}_{{tL}}(x) \left({   }\partial
   _x\uppsi  ^1{}_{{tL}}(x) \right)+{\sin }\left[p^1{}_{{tR}}(x)-p^2{}_{{tR}}(x)\right] \uppsi
   ^2{}_{{tR}}(x) \left({   }\partial _x\uppsi  ^1{}_{{tR}}(x) \right)-2
   {\sin }\left[p^1{}_{{bR}}(x)-p^2{}_{{bR}}(x)\right] \uppsi  ^1{}_{{bR}}(x) \left({   }\partial _x\uppsi
   ^2{}_{{bR}}(x) \right)-\uppsi  ^1{}_{{tL}}(x)
   \left({\cos }\left[p^1{}_{{tL}}(x)-p^2{}_{{tL}}(x)\right] \left({   }\partial _y\uppsi  ^2{}_{{tL}}(x)
   \right)+\uppsi  ^2{}_{{tL}}(x) \left({\sin }\left[p^1{}_{{tL}}(x)-p^2{}_{{tL}}(x)\right] \left({
   }\partial _yp^1{}_{{tL}}(x) \right)+{\sin }\left[p^1{}_{{tL}}(x)-p^2{}_{{tL}}(x)\right] \left({
   }\partial _yp^2{}_{{tL}}(x) \right)-{\cos }\left[p^1{}_{{tL}}(x)-p^2{}_{{tL}}(x)\right] \left({
   }\partial _xp^1{}_{{tL}}(x) +{   }\partial _xp^2{}_{{tL}}(x)
   \right)\right)+{\sin }\left[p^1{}_{{tL}}(x)-p^2{}_{{tL}}(x)\right] \left({   }\partial _x\uppsi
   ^2{}_{{tL}}(x) \right)\right)-\uppsi  ^1{}_{{tR}}(x)
   \left({\cos }\left[p^1{}_{{tR}}(x)-p^2{}_{{tR}}(x)\right] \left({   }\partial _y\uppsi  ^2{}_{{tR}}(x)
   \right)+\uppsi  ^2{}_{{tR}}(x) \left({\sin }\left[p^1{}_{{tR}}(x)-p^2{}_{{tR}}(x)\right] \left({
   }\partial _yp^1{}_{{tR}}(x) \right)+{\sin }\left[p^1{}_{{tR}}(x)-p^2{}_{{tR}}(x)\right] \left({
   }\partial _yp^2{}_{{tR}}(x) \right)-{\cos }\left[p^1{}_{{tR}}(x)-p^2{}_{{tR}}(x)\right] \left({
   }\partial _xp^1{}_{{tR}}(x) +{   }\partial _xp^2{}_{{tR}}(x)
   \right)\right)+{\sin }\left[p^1{}_{{tR}}(x)-p^2{}_{{tR}}(x)\right] \left({   }\partial _x\uppsi
   ^2{}_{{tR}}(x) \right)\right)-2 \uppsi  ^1{}_{{bR}}(x){}^2 \left({   }\partial _tp^1{}_{{bR}}(x) \right)-\uppsi
   ^1{}_{{tL}}(x){}^2 \left(-\left({   }\partial _zp^1{}_{{tL}}(x) \right)+{   }\partial
   _tp^1{}_{{tL}}(x) \right)-\uppsi  ^1{}_{{tR}}(x){}^2 \left({   }\partial _zp^1{}_{{tR}}(x) +{   }\partial
   _tp^1{}_{{tR}}(x) \right)-2 \uppsi  ^2{}_{{bR}}(x){}^2 \left({   }\partial _tp^2{}_{{bR}}(x) \right)-\uppsi
   ^2{}_{{tL}}(x){}^2 \left({   }\partial _tp^2{}_{{tL}}(x) \right)-\uppsi  ^2{}_{{tR}}(x){}^2 \left({
   }\partial _tp^2{}_{{tR}}(x) \right).
   \]


\subsection {Yukawa. $\frac{\sqrt{2}}{v}  [ m_t   { \bar   t_R      }(x)    {\tilde   {\bf H}^\dagger  (x)} {  \bf q }_L(x) +
 m_b \bar{
  \bf q }_L(x)  {\bf H}(x)   b_R  (x) ]    +\{ hc \}
\leftrightarrow   {\rm tr}
   \frac{\sqrt{2}}{v}  [ m_t   \Psi_{tR} ^\dagger(x) {\bf   H}_{t} (x) { \boldmathPsi}_{qL}(x) +
 m_b {  \boldmathPsi}_{qL}  ^\dagger(x) {\bf  H}_{b}(x) \Psi_{bR} (x)]    +\{ hc \} $}

 The representation of scalars in the conventional and spin  bases  uses    the  association, e. g.,
  ${\bf H}   {\gamma ^0} _{4\times4} \rightarrow {\bf  H}_t $;  the conventional  phases, written explicitly
   in Appendix 2, are set to fit the spin basis,   as both operators act
  equally on fermions, and we  applied the   gamma-matrix  representation freedom of choice.

 \[
\frac{1}{v}\left[ \cos \left(p_{{W2}}-p_{{Z1}}-p_{{\eta 1}}(x)-p^1{}_{{bR}}(x)+p^1{}_{{tL}}(x)\right) m_b \eta ^r{}_1(x) \uppsi
   ^1{}_{{bR}}(x) \uppsi ^1{}_{{tL}}(x)+\cos \left(p_{{W2}}-p_{{Z2}}-p_{{\eta 0}}(x)+p^1{}_{{tL}}(x)-p^1{}_{{tR}}(x)\right) m_t \eta ^r{}_0(x) \uppsi ^1{}_{{tR}}(x) \uppsi ^1{}_{{tL}}(x)+\uppsi
   ^1{}_{{bL}}(x) \left(\sin \left(p_{{W2}}-p_{{Z2}}+p_{{\eta 1}}(x)+p^1{}_{{bL}}(x)-p^1{}_{{tR}}(x)\right)
  m_t \eta ^r{}_1(x) \uppsi ^1{}_{{tR}}(x)-\sin \left(p_{{W2}}-p_{{Z1}}+p_{{\eta 0}}(x)+p^1{}_{{bL}}(x)-p^1{}_{{bR}}(x)\right) m_b \eta ^r{}_0(x) \uppsi ^1{}_{{bR}}(x)\right)-\cos
   \left(p_{{W2}}-p_{{Z1}}-p_{{\eta 1}}(x)-p^2{}_{{bR}}(x)+p^2{}_{{tL}}(x)\right) m_b \eta ^r{}_1(x) \uppsi
   ^2{}_{{bR}}(x) \uppsi ^2{}_{{tL}}(x)-\cos \left(p_{{W2}}-p_{{Z2}}-p_{{\eta 0}}(x)+p^2{}_{{tL}}(x)-p^2{}_{{tR}}(x)\right) m_t \eta ^r{}_0(x) \uppsi ^2{}_{{tL}}(x) \uppsi ^2{}_{{tR}}(x)+\uppsi
   ^2{}_{{bL}}(x) \left(\sin \left(p_{{W2}}-p_{{Z1}}+p_{{\eta 0}}(x)+p^2{}_{{bL}}(x)-p^2{}_{{bR}}(x)\right)
   m_b \eta ^r{}_0(x) \uppsi ^2{}_{{bR}}(x)-\sin \left(p_{{W2}}-p_{{Z2}}+p_{{\eta 1}}(x)+p^2{}_{{bL}}(x)-p^2{}_{{tR}}(x)\right) m_t \eta ^r{}_1(x) \uppsi ^2{}_{{tR}}(x)\right) \right ],
\]
requiring the identities
$ p_{  {Z1}}-\pi=  p_{  {Z2}}- \frac{\pi}{2}= p_{  {W1}}$.

\section*{Appendix 2: Scalar-vector Lagrangian  ${\cal L}_{SV}$;  conjugate-Higgs invariance  }

For the scalar components, we also use  expressions in polar coordinates, and in which the phase is written explicitly, to see its workings.  Thus,
for the conventional basis,

\[{\bf H}(x)=\frac{1}{\sqrt{2}} \left(\begin{array}{c}
  \eta_{1}^r(x) e^{ i p^1{}_{t}+i p_{\eta  1}(x)} \\
   \eta_0^r(x)e^{ i p^0{}_{t}+i p_{\eta 0}(x)}
  \end{array}\right) \]

 \[{\tilde {\bf H}}(x)=\frac{1}{\sqrt{2}} \left(\begin{array}{c}
  -i\eta_0^r(x) e^{ i p^0{}_{b}-i p_{\eta 0}(x)}  \\
  i\eta_{1}^r(x) e^{i p^1{}_{b}-i p_{\eta  1}(x)}
  \end{array}\right) \]
where $p^1{}_{b}, p^1{}_{t}, p^0{}_{b}, p^0{}_{t}$  are charged and neutral phases, respectively, and
 $\bar{\bf H}_{ \chi_t ,  \chi_b }(x)=(\chi_t  {\bf H}(x),\chi_b \tilde{\bf H}(x))$  is a $4 \times 4$
matrix,
 $\chi_t,$  $\chi_b,$ can be assumed real and their dependence  in all terms is through the factor  $\chi_t^2+\chi_b^2$, so their explicit form constitutes a  likewise  demonstration for ${\cal L}_{SV}$.

 For the spin  basis, we use a generalized expression for the scalar term with conjugated terms weighted by a  multiplicative parameter $\lambda$,  to keep track of terms, and with a normalization that makes  ${\cal L}_{SV}$ $\lambda$-independent:  \[ \label{phibolddef}
{\bf H}_{ab}^{tot\lambda }(x)  =   \frac{1}{\sqrt{(1+\lambda^2}}
 \left [ \chi_t  \left [ \eta_{1}^r(x) e^{ i \phi^{1}{}_t+i p_{\eta  1}(x)}   ( \phi_{1}^{+} +\phi_{2}^{+})+
      \eta_{0}^r(x)
 e^{i  \phi^{0}{}_t+i p_{\eta  0}(x) }  ( \phi_{1}^{0} +\phi_{2}^{0}) \right ]+\chi_b \left [  \eta_{1}^r(x) e^{ i \phi^{1}{}_b-i p_{\eta    1}(x)}   ( \phi_{1}^{+}-\phi_{2}^{+}) ^ \dagger
   +
      \eta_{0}^r(x)  e^{i  \phi^{0}{}_b-i p_{\eta  0}(x) }   ( \phi_{1}^{0} -\phi_{2}^{0}) ^ \dagger
     \right ]+
 \lambda  \chi_t \left [  \eta_{1}^r(x)       e^{ i \phi^{1}{}_{\lambda t}-i p_{\eta  1}(x)}
   ( \phi_{1}^{+} +\phi_{2}^{+}) ^ \dagger+
   \eta_{0}^r(x)   e^{i  \phi^{0}{}_{\lambda t}-i p_{\eta  0}(x) }   ( \phi_{1}^{0} +\phi_{2}^{0})  ^ \dagger  \right ]
    +
 \lambda \chi_b   \left [  \eta_{1}^r(x)  e^{i  \phi^{1}{}_{\lambda b}+i p_{\eta  1}(x) }
   ( \phi_{1}^{+} -\phi_{2}^{+})+
  \eta_{0}^r(x)  e^{i  \phi^{0}{}_{\lambda b}+i p_{\eta  0}(x) }  ( \phi_{1}^{0} -\phi_{2}^{0}) \right ]
    \right ]
 , \]
where $\phi_{1,2}^{+}$, $\phi_{1,2}^{0}$ are defined in Table 1, and
$ \phi^{  1}{}_{t}, \phi^{  1}{}_{b}, \phi^{  0}{}_{t}, \phi^{  0}{}_{b}$ are charged and neutral phases, respectively, and those with $\lambda$ correspond to the hermitian-conjugate function  (see Eqs. (50), (51) in the paper.)
Given the chiral
 nature of the scalar components, they do not mix with their hermitian-conjugate components.

 Thus, ${\cal L}_{SV}= {\bf H}^\dagger(x) {{\bf F}^\mu}^\dagger (x)  {\bf F}_\mu(x) {\bf H}(x)$, with ${\bf F}_\mu(x)= i \partial_\mu+\frac{1}{2}g  {\boldmathtau}\cdot {{\bf W}_\mu}(x)+\frac{1}{2}g'B_\mu(x) $, 
 ${{\bf W}_\mu}(x)=(W^1_\mu(x),W^2_\mu(x),W^3_\mu(x))$  (cf. Eqs. (42) and (43)) is  compared with
 
\noindent $\frac{1}{2} { \rm tr  }\{[{\bf F''}(x) ,{\bf H}_{ab} ^{tot\lambda}(x) ]_\pm^\dagger
[ {\bf F''}(x) ,{\bf H}_{ab} ^{tot\lambda}(x)  ]_\pm \}_{\rm sym} $,    where  ${\bf F''}(x)=
[i\partial_\mu+    {  g W_\mu^i(x) I_i} +\frac{1}{2}g' B_\mu(x) Y_o]\gamma_0 \gamma^\mu $; the subindex {\it sym}  means only symmetric $\gamma_\mu \gamma_\nu$  components are taken.


\subsection {Square W.   ${ \rm tr  }{\bf \bar  H}^\dagger(x) \frac{1}{2}g  {\boldmathtau}\cdot {{\bf W}_\mu}(x) \frac{1}{2}g  {\boldmathtau}\cdot {{\bf W}^\mu}(x){\bf \bar H}(x)
\leftrightarrow  \\ \frac{1}{2} { \rm tr  }
   \{ [ {  g W_0^n(x) I_n} \gamma_0 \gamma^0 ,{\bf H}_{ab}^{tot\lambda }(x) ]^\dagger
[ {  g W_0^m(x) I_m} \gamma_0 \gamma^0 ,{\bf H}_{ab}^{tot\lambda }(x) ]+ \\  \{{  g W_j^n(x) I_n} \gamma_0 \gamma^j ,{\bf H}_{ab}^{tot\lambda }(x) \}^\dagger
\{ {  g W_k^m(x) I_m} \gamma_0 \gamma^k ,{\bf H}_{ab}^{tot\lambda }(x) \} \} $ }
\[
 \frac{1}{8}{g}^2  \left(\chi _t^2+\chi _{ b}^2\right) \left( \eta _0^r{} (x) ^2+\eta _1^r(x){}^2\right) W_{\mu }^n(x) W^{n \mu }  (x)
\]

\subsection {Square B. ${ \rm tr  }{\bf \bar H}^\dagger(x)\frac{1}{2}g'B_\mu(x)\frac{1}{2}g'B^\mu(x){\bf \bar H}(x)
\leftrightarrow \\ \frac{1}{2} { \rm tr  }
   \{ [ {  \frac{1}{2}g' B_0(x) Y_o} \gamma_0 \gamma^0 ,{\bf H}_{ab}^{tot\lambda }(x) ]^\dagger
[ {  \frac{1}{2}g' B_0(x) Y_o} \gamma_0 \gamma^0 ,{\bf H}_{ab}^{tot\lambda }(x) ]+   \\ \{{  \frac{1}{2}g' B_j(x) Y_o} \gamma_0 \gamma^j ,{\bf H}_{ab}^{tot\lambda }(x) \}^\dagger
\{ {  \frac{1}{2}g' B_k(x) Y_o} \gamma_0 \gamma^k ,{\bf H}_{ab}^{tot\lambda }(x) \} \}    $ }
\[
 \frac{1}{8}{g'}^2 \left(\chi_t^2+\chi_b^2\right)  \left(  \eta_0 ^r{}(x) ^2+\eta _1^r(x){}^2\right) B_{\mu }(x) B^{\mu }(x)
\]

\subsection {Cross B-W. ${ \rm tr  }\{ {\bf \bar H}^\dagger(x)     \frac{1}{2}g  {\boldmathtau}\cdot {{\bf W}_\mu}(x)
 \frac{1}{2}g'B^\mu(x) {\bf \bar  H}(x)+\\ {\bf \bar H}^\dagger(x)
 \frac{1}{2}g'B^\mu(x)      \frac{1}{2}g  {\boldmathtau}\cdot {{\bf W}_\mu}(x)
 {\bf \bar  H}(x) \}
\leftrightarrow \\ \frac{1}{2} { \rm tr  }
   \{  [ {  \frac{1}{2}g' B_0(x) Y_o} \gamma_0 \gamma^0 ,{\bf H}_{ab}^{tot\lambda }(x) ]^\dagger
[  g W_0^n(x) I_n \gamma_0 \gamma^0 ,{\bf H}_{ab}^{tot\lambda }(x) ]\\ +
 [ g W_0^n(x) I_n \gamma_0 \gamma^0  ,{\bf H}_{ab}^{tot\lambda }(x) ]^\dagger
[  {  \frac{1}{2}g' B_0(x) Y_o} \gamma_0 \gamma^0 ,{\bf H}_{ab}^{tot\lambda }(x) ]  \\ +
   \{{\frac{1}{2}g' B_j(x) Y_o} \gamma_0 \gamma^j ,{\bf H}_{ab}^{tot\lambda }(x) \}^\dagger
\{ { g W_k^n(x) I_n} \gamma_0 \gamma^k ,{\bf H}_{ab}^{tot\lambda }(x) \} \\  +
    \{{ g W_j^n(x) I_n} \gamma_0 \gamma^j ,{\bf H}_{ab}^{tot\lambda }(x) \}^\dagger
\{ {\frac{1}{2}g' B_k(x) Y_o} \gamma_0 \gamma^k ,{\bf H}_{ab}^{tot\lambda }(x) \}  \} $ }

\[ \frac{1}{4} g {g'} \left(\chi _t^2+\chi _{b}^2\right) \left(-B_0(x) W^3{}_0(x) \eta
   ^r{}_0(x){}^2+B_1(x) W^3{}_1(x) \eta ^r{}_0(x){}^2+B_2(x) W^3{}_2(x) \eta ^r{}_0(x){}^2+B_3(x)
   W^3{}_3(x) \eta ^r{}_0(x){}^2+2  {\cos  }\left[p^0{}_t-p^1{}_b-p_{ {\eta 0}}(x)+p_{ {\eta 1}}(x)\right] B_0(x) W^1{}_0(x) \eta ^r{}_0(x) \eta ^r{}_1(x)-2
    {\sin }\left[p^0{}_t-p^1{}_b-p_{ {\eta 0}}(x)+p_{ {\eta 1}}(x)\right] B_0(x)
   W^2{}_0(x) \eta ^r{}_0(x) \eta ^r{}_1(x)-2  {\cos  }\left[p^0{}_t-p^1{}_b-p_{ {\eta 0}}(x)+p_{ {\eta 1}}(x)\right] B_1(x) W^1{}_1(x) \eta ^r{}_0(x) \eta ^r{}_1(x)+2
    {\sin }\left[p^0{}_t-p^1{}_b-p_{ {\eta 0}}(x)+p_{ {\eta 1}}(x)\right] B_1(x)
   W^2{}_1(x) \eta ^r{}_0(x) \eta ^r{}_1(x)-2  {\cos  }\left[p^0{}_t-p^1{}_b-p_{ {\eta 0}}(x)+p_{ {\eta 1}}(x)\right] B_2(x) W^1{}_2(x) \eta ^r{}_0(x) \eta ^r{}_1(x)+2
    {\sin }\left[p^0{}_t-p^1{}_b-p_{ {\eta 0}}(x)+p_{ {\eta 1}}(x)\right] B_2(x)
   W^2{}_2(x) \eta ^r{}_0(x) \eta ^r{}_1(x)-2  {\cos  }\left[p^0{}_t-p^1{}_b-p_{ {\eta 0}}(x)+p_{ {\eta 1}}(x)\right] B_3(x) W^1{}_3(x) \eta ^r{}_0(x) \eta ^r{}_1(x)+2
    {\sin }\left[p^0{}_t-p^1{}_b-p_{ {\eta 0}}(x)+p_{ {\eta 1}}(x)\right] B_3(x)
   W^2{}_3(x) \eta ^r{}_0(x) \eta ^r{}_1(x)+B_0(x) W^3{}_0(x) \eta ^r{}_1(x){}^2-B_1(x)
   W^3{}_1(x) \eta ^r{}_1(x){}^2-B_2(x) W^3{}_2(x) \eta ^r{}_1(x){}^2-B_3(x) W^3{}_3(x) \eta
   ^r{}_1(x){}^2\right) . \]

\subsection {Cross W-derivative. ${ \rm tr  }\{{\bf \bar H}^\dagger(x)     \frac{1}{2}g  {\boldmathtau}\cdot {{\bf W}_\mu}(x)
i \partial ^ \mu {\bf \bar  H}(x)-\\ {\bf \bar H}^\dagger(x)
 i \overleftarrow{ \partial ^ \mu}    \frac{1}{2}g  {\boldmathtau}\cdot {{\bf W}_\mu}(x)
 {\bf \bar  H}(x) \}
\leftrightarrow \\ \frac{1}{2} { \rm tr  }
   \{  [ {   -i \overleftarrow{\partial _0} }\gamma_0 \gamma^0,{\bf H}_{ab}^{tot\lambda }(x) ]^\dagger
[  g W_0^n(x) I_n \gamma_0 \gamma^0 ,{\bf H}_{ab}^{tot\lambda }(x) ] \\    +
   [ g W_0^n(x) I_n \gamma_0 \gamma^0  ,{\bf H}_{ab}^{tot\lambda }(x) ]^\dagger
[ i  \partial _ 0  \gamma_0 \gamma^0 ,{\bf H}_{ab}^{tot\lambda }(x) ]+   \\
   \{-i \overleftarrow{\partial_j} \gamma_0 \gamma^j ,{\bf H}_{ab}^{tot\lambda }(x) \}^\dagger
\{ { g W_k^n(x) I_n} \gamma_0 \gamma^k ,{\bf H}_{ab}^{tot\lambda }(x) \}+ \\
    \{{ g W_j^n(x) I_n} \gamma_0 \gamma^j ,{\bf H}_{ab}^{tot\lambda }(x) \}^\dagger
\{ i \partial _k  \gamma_0 \gamma^k,{\bf H}_{ab}^{tot\lambda }(x) \}  \} $ }
 As for the kinetic term in  ${\cal L}_{FV}$, the fields' integrability property leads to derivatives in the form  $ i\frac{1}{2}\stackrel{\leftrightarrow}{\partial_\mu}$; similarly   for the cross B-derivative and d'Alembert terms next.


\[
\frac{1}{2}g\left(\chi _t^2+\chi _{b}^2\right) \left(\eta ^r{}_1(x){}^2 \left(W^3{}_3(x) \left({   }\partial _zp_{{\eta 1}}(x) \right)+W^3{}_2(x) \left({   }\partial _yp_{{\eta 1}}(x)
   \right)+W^3{}_1(x) \left({   }\partial _xp_{{\eta 1}}(x) \right)-W^3{}_0(x) \left({   }\partial _tp_{{\eta 1}}(x) \right)\right)-\eta ^r{}_1(x) \left(W^2{}_3(x)
   \left({{\rm sin}}\left[p^0{}_b-p^1{}_b-p_{{\eta 0}}(x)+p_{{\eta 1}}(x)\right] \eta ^r{}_0(x) \left({   }\partial _zp_{{\eta 0}}(x) +{   }\partial _zp_{{\eta 1}}(x)
   \right)\\ \nonumber
   +{{\rm cos}}\left[p^0{}_b-p^1{}_b-p_{{\eta 0}}(x)+p_{{\eta 1}}(x)\right] \left({   }\partial _z\eta ^r{}_0(x) \right)\right)+W^1{}_3(x) \left(-{{\rm cos}}\left[p^0{}_b-p^1{}_b-p_{{\eta 0}}(x)+p_{{\eta 1}}(x)\right] \eta ^r{}_0(x) \left({   }\partial _zp_{{\eta 0}}(x) +{   }\partial _zp_{{\eta 1}}(x) \right) \\ \nonumber
   +{{\rm sin}}\left[p^0{}_b-p^1{}_b-p_{{\eta 0}}(x)+p_{{\eta 1}}(x)\right] \left({   }\partial _z\eta ^r{}_0(x) \right)\right)-{{\rm cos}}\left[p^0{}_b-p^1{}_b-p_{{\eta 0}}(x)+p_{{\eta 1}}(x)\right] W^1{}_2(x) \eta ^r{}_0(x)
   \left({   }\partial _yp_{{\eta 0}}(x) \right)+{{\rm sin}}\left[p^0{}_b-p^1{}_b-p_{{\eta 0}}(x)+p_{{\eta 1}}(x)\right] W^2{}_2(x) \eta ^r{}_0(x) \left({   }\partial _yp_{{\eta 0}}(x) \right)-{{\rm cos}}\left[p^0{}_b-p^1{}_b-p_{{\eta 0}}(x)+p_{{\eta 1}}(x)\right] W^1{}_2(x) \eta ^r{}_0(x) \left({   }\partial _yp_{{\eta 1}}(x)
   \right)+{{\rm sin}}\left[p^0{}_b-p^1{}_b-p_{{\eta 0}}(x)+p_{{\eta 1}}(x)\right] W^2{}_2(x) \eta ^r{}_0(x) \left({   }\partial _yp_{{\eta 1}}(x) \right)+{{\rm sin}}\left[p^0{}_b-p^1{}_b-p_{{\eta 0}}(x)+p_{{\eta 1}}(x)\right] W^1{}_2(x) \left({   }\partial _y\eta ^r{}_0(x) \right)\\ \nonumber
   +{{\rm cos}}\left[p^0{}_b-p^1{}_b-p_{{\eta 0}}(x)+p_{{\eta 1}}(x)\right] W^2{}_2(x) \left({
   }\partial _y\eta ^r{}_0(x) \right)-{{\rm cos}}\left[p^0{}_b-p^1{}_b-p_{{\eta 0}}(x)+p_{{\eta 1}}(x)\right] W^1{}_1(x) \eta ^r{}_0(x) \left({   }\partial _xp_{{\eta 0}}(x)
   \right)\\ \nonumber
   +{{\rm sin}}\left[p^0{}_b-p^1{}_b-p_{{\eta 0}}(x)+p_{{\eta 1}}(x)\right] W^2{}_1(x) \eta ^r{}_0(x) \left({   }\partial _xp_{{\eta 0}}(x) \right)-{{\rm cos}}\left[p^0{}_b-p^1{}_b-p_{{\eta 0}}(x)+p_{{\eta 1}}(x)\right] W^1{}_1(x) \eta ^r{}_0(x) \left({   }\partial _xp_{{\eta 1}}(x) \right)\\ \nonumber
   +{{\rm sin}}\left[p^0{}_b-p^1{}_b-p_{{\eta 0}}(x)+p_{{\eta 1}}(x)\right]
   W^2{}_1(x) \eta ^r{}_0(x) \left({   }\partial _xp_{{\eta 1}}(x) \right)+{{\rm sin}}\left[p^0{}_b-p^1{}_b-p_{{\eta 0}}(x)+p_{{\eta 1}}(x)\right] W^1{}_1(x) \left({   }\partial _x\eta
   ^r{}_0(x) \right)\\
   +{{\rm cos}}\left[p^0{}_b-p^1{}_b-p_{{\eta 0}}(x)+p_{{\eta 1}}(x)\right] W^2{}_1(x) \left({   }\partial _x\eta ^r{}_0(x) \right)\\
   +{{\rm cos}}\left[p^0{}_b-p^1{}_b-p_{{\eta 0}}(x)+p_{{\eta 1}}(x)\right] W^1{}_0(x) \eta ^r{}_0(x) \left({   }\partial _tp_{{\eta 0}}(x) \right)\\
   -{{\rm sin}}\left[p^0{}_b-p^1{}_b-p_{{\eta 0}}(x)+p_{{\eta 1}}(x)\right]
   W^2{}_0(x) \eta ^r{}_0(x) \left({   }\partial _tp_{{\eta 0}}(x) \right)\\
   +{{\rm cos}}\left[p^0{}_b-p^1{}_b-p_{{\eta 0}}(x)+p_{{\eta 1}}(x)\right] W^1{}_0(x) \eta ^r{}_0(x) \left({
   }\partial _tp_{{\eta 1}}(x) \right)\\
   -{{\rm sin}}\left[p^0{}_b-p^1{}_b-p_{{\eta 0}}(x)+p_{{\eta 1}}(x)\right] W^2{}_0(x) \eta ^r{}_0(x) \left({   }\partial _tp_{{\eta 1}}(x)
   \right)\\
   -{{\rm sin}}\left[p^0{}_b-p^1{}_b-p_{{\eta 0}}(x)+p_{{\eta 1}}(x)\right] W^1{}_0(x) \left({   }\partial _t\eta ^r{}_0(x) \right)\\
   -{{\rm cos}}\left[p^0{}_b-p^1{}_b-p_{{\eta 0}}(x)+p_{{\eta 1}}(x)\right] W^2{}_0(x) \left({   }\partial _t\eta ^r{}_0(x) \right)\right)+\eta ^r{}_0(x) \left(-W^3{}_3(x) \eta ^r{}_0(x) \left({   }\partial _zp_{{\eta 0}}(x)
   \right)+{{\rm sin}}\left[p^0{}_b-p^1{}_b-p_{{\eta 0}}(x)+p_{{\eta 1}}(x)\right] W^1{}_3(x) \left({   }\partial _z\eta ^r{}_1(x) \right)+{{\rm cos}}\left[p^0{}_b-p^1{}_b-p_{{\eta 0}}(x)+p_{{\eta 1}}(x)\right] W^2{}_3(x) \left({   }\partial _z\eta ^r{}_1(x) \right)-W^3{}_2(x) \eta ^r{}_0(x) \left({   }\partial _yp_{{\eta 0}}(x) \right)\\
   +{{\rm sin}}\left[p^0{}_b-p^1{}_b-p_{{\eta 0}}(x)+p_{{\eta 1}}(x)\right] W^1{}_2(x) \left({   }\partial _y\eta ^r{}_1(x) \right)\\
   +{{\rm cos}}\left[p^0{}_b-p^1{}_b-p_{{\eta 0}}(x)+p_{{\eta 1}}(x)\right] W^2{}_2(x) \left({
   }\partial _y\eta ^r{}_1(x) \right)-W^3{}_1(x) \eta ^r{}_0(x) \left({   }\partial _xp_{{\eta 0}}(x) \right)\\
   +{{\rm sin}}\left[p^0{}_b-p^1{}_b-p_{{\eta 0}}(x)+p_{{\eta 1}}(x)\right] W^1{}_1(x)
   \left({   }\partial _x\eta ^r{}_1(x) \right)\\
   +{{\rm cos}}\left[p^0{}_b-p^1{}_b-p_{{\eta 0}}(x)+p_{{\eta 1}}(x)\right] W^2{}_1(x) \left({   }\partial _x\eta ^r{}_1(x) \right)+W^3{}_0(x) \eta
   ^r{}_0(x) \left({   }\partial _tp_{{\eta 0}}(x) \right)\\
   -{{\rm sin}}\left[p^0{}_b-p^1{}_b-p_{{\eta 0}}(x)+p_{{\eta 1}}(x)\right] W^1{}_0(x) \left({   }\partial _t\eta ^r{}_1(x)
   \right)\\
   -{{\rm cos}}\left[p^0{}_b-p^1{}_b-p_{{\eta 0}}(x)+p_{{\eta 1}}(x)\right] W^2{}_0(x) \left({   }\partial _t\eta ^r{}_1(x) \right)\right)\right),
\]
from which one derives the phase connections

\[ \phi ^1{}_{{\lambda t}}= p^0{}_t-p^1{}_t+\phi ^0{}_{{\lambda t}}-\frac{\pi
   }{2} \]
   \[ \phi ^1{}_b= p^0{}_t-p^1{}_t+\phi ^0{}_b-\frac{\pi }{2} \]

\[ \phi ^1{}_t= p^0{}_b-p^1{}_b+\phi ^0{}_t+\frac{\pi }{2} \]   \[ \phi ^1{}_{{ \lambda  b}}=
   p^0{}_b-p^1{}_b+\phi ^0{}_{{ \lambda  b}}+\frac{\pi }{2}. \]

\subsection {Cross B-derivative. ${ \rm tr  }\{{\bf \bar H}^\dagger(x)    \frac{1}{2}g'B_\mu(x)
i \partial ^ \mu {\bf \bar  H}(x)-\\ {\bf \bar H}^\dagger(x)
 i \overleftarrow{ \partial ^ \mu}    \frac{1}{2}g'B_\mu(x)
 {\bf \bar  H}(x) \}
\leftrightarrow \\ \frac{1}{2} { \rm tr  }
   \{  [ {   -i \overleftarrow{\partial _0} }\gamma_0 \gamma^0,{\bf H}_{ab}^{tot\lambda }(x) ]^\dagger
[ {\frac{1}{2}g' B_0(x) Y_o} \gamma_0 \gamma^0 ,{\bf H}_{ab}^{tot\lambda }(x) ]\\ +
  [ {\frac{1}{2}g' B_0(x) Y_o} \gamma_0 \gamma^0 ,{\bf H}_{ab}^{tot\lambda }(x) ]^\dagger
[ i  \partial ^ 0  \gamma_0 \gamma^0 ,{\bf H}_{ab}^{tot\lambda }(x) ]+ \\
   \{-i \overleftarrow{\partial _j} \gamma_0 \gamma^j ,{\bf H}_{ab}^{tot\lambda }(x) \}^\dagger
\{ {\frac{1}{2}g' B_k(x) Y_o} \gamma_0 \gamma^k ,{\bf H}_{ab}^{tot\lambda }(x) \}+ \\
    \{{\frac{1}{2}g' B_j(x) Y_o} \gamma_0 \gamma^j ,{\bf H}_{ab}^{tot\lambda }(x) \}^\dagger
\{ i \partial _k  \gamma_0 \gamma^k,{\bf H}_{ab}^{tot\lambda }(x) \}  \}. $ }

\[
\frac{1}{2}  \left(\chi _b^2+\chi _t^2\right) g' \left(\left(B_3(x) \left({   }\partial _zp_{{\eta 0}}(x)\right)+B_2(x) \left({   }\partial _yp_{{\eta 0}}(x)\right)+B_1(x)
   \left({   }\partial _xp_{{\eta 0}}(x)\right)-B_0(x) \left({   }\partial
   _tp_{{\eta 0}}(x)\right)\right) \eta ^r{}_0(x){}^2+\left(B_3(x) \left({
   }\partial _zp_{{\eta 1}}(x)\right)+B_2(x) \left({   }\partial _yp_{{\eta 1}}(x)\right)+B_1(x) \left({   }\partial _xp_{{\eta 1}}(x)\right)-B_0(x)
   \left({   }\partial _tp_{{\eta 1}}(x)\right)\right) \eta ^r{}_1(x){}^2\right).
\]
In addition to the above equations, we derive
\[p^0{}_t= -p^0{}_b+p^1{}_b+p^1{}_t \].
  As the similarity transformation phases in e.g. ${\bf H}$, $\tilde {\bf H}$, this relation accounts for the sign change for complex conjugate components.

\subsection {d'Alembert. $  { \rm tr  }  { \bf \bar H}^\dagger(x)    \overleftarrow{ \partial ^ \mu}
  \partial_\mu{\bf \bar H}(x)
\leftrightarrow \\ -\frac{1}{2} { \rm tr  }
   \{ [ {    \overleftarrow{\partial _0}} \gamma_0 \gamma^0 ,{\bf H}_{ab}^{tot\lambda }(x) ]^\dagger
[ {   \partial _0} \gamma_0 \gamma^0 ,{\bf H}_{ab}^{tot\lambda }(x) ]-  \\ \{{ \overleftarrow{\partial _j} }\gamma_0 \gamma^j ,{\bf H}_{ab}^{tot\lambda }(x) \}^\dagger
\{ {  \partial _k}  \gamma_0 \gamma^k ,{\bf H}_{ab}^{tot\lambda }(x) \} \} $ }
\[
-\frac{1}{2} \left(\chi _b^2+\chi _t^2\right) \left(\left({   }\partial _z\eta
   _1^r(x) \right){}^2+\eta ^r{}_0(x){}^2 \left({   }\partial
   _zp_{{\eta 0}}(x) \right){}^2+\eta _1^r(x){}^2 \left({   }\partial
   _zp_{{\eta 1}}(x) \right){}^2+\left({   }\partial _z\eta ^r{}_0(x)
   \right){}^2+\left({   }\partial _y\eta _1^r(x) \right){}^2+\eta
   ^r{}_0(x){}^2 \left({   }\partial _yp_{{\eta 0}}(x)
   \right){}^2+\eta _1^r(x){}^2 \left({   }\partial _yp_{{\eta 1}}(x)
   \right){}^2+\left({   }\partial _y\eta ^r{}_0(x) \right){}^2+\left({
   }\partial _x\eta _1^r(x) \right){}^2+\eta ^r{}_0(x){}^2 \left({
   }\partial _xp_{{\eta 0}}(x) \right){}^2+\eta _1^r(x){}^2 \left({
   }\partial _xp_{{\eta 1}}(x) \right){}^2+\left({   }\partial _x\eta
   ^r{}_0(x) \right){}^2-\left({   }\partial _t\eta _1^r(x) \right){}^2-\eta
   ^r{}_0(x){}^2 \left({   }\partial _tp_{{\eta 0}}(x)
   \right){}^2-\eta _1^r(x){}^2 \left({   }\partial _tp_{{\eta 1}}(x)
   \right){}^2-\left({   }\partial _t\eta ^r{}_0(x) \right){}^2\right).
   \]

Each of the  ${\cal L}_{SV}$    terms is indeed proportional to the combination $\chi _b^2+\chi _t^2 $, which manifests
the t-b symmetry of this component, as the phases that connect the two representations were obtained.

We thus completed the demonstration of the SM Lagrangian terms'  equivalence
 in two bases; we conclude the spin-space representation  reproduces the same properties of   SM generators.












\end{document}